\documentclass[a4paper,11pt]{article}

\pdfoutput=1
\usepackage{amsfonts}

\usepackage{graphicx}
\usepackage{amsmath}
\usepackage{amssymb}
\usepackage{latexsym}
\usepackage[colorlinks]{hyperref}
\usepackage{color}
\usepackage{float}
\usepackage{cite}
\usepackage{makeidx}
\usepackage{colortbl}
\usepackage{csquotes}
\usepackage[squaren]{SIunits}

\usepackage[textfont={footnotesize,sf},labelfont={color=blue,bf,sf},labelsep=endash]{caption}
\usepackage[position=top,labelfont={color=blue,bf,sf}]{subfig}
\usepackage{bm}

\newcommand{\bra}{\begin{array}}
\newcommand{\era}{\end{array}}
\newcommand{\beq}{\begin{equation}}
\newcommand{\eeq}{\end{equation}}
\newcommand{\bqr}{\begin{eqnarray}}
\newcommand{\eqr}{\end{eqnarray}}

\def\BC{\bb C}
\def\_\BC{\bbi C}

\def\Tr {{\rm Tr}}

\def\( {\left(}
\def\) {\right)}
\def\no2 {{\textstyle{n\over 2}}}

\def\Tr {{\rm Tr}}

\newcommand{\pa}{\partial}

\newcommand{\del}{\delta}
\newcommand{\lga}{\longrightarrow}

\newcommand{\lb}{\label}
\newcommand{\nn}{\nonumber}

\begin{document}

\begin{titlepage}
\setcounter{page}{1}
\renewcommand{\thefootnote}{\fnsymbol{footnote}}

\begin{flushright}
\end{flushright}

\vspace{5mm}
\begin{center}

{\Large \bf {Entropies for  Coupled  Harmonic Oscillators and Temperature}}

\vspace{5mm}

 {\bf Ahmed Jellal\footnote{\sf 
a.jellal@ucd.ac.ma}}$^{a}$
and
{\bf Abdeldjalil Merdaci\footnote{\sf amerdaci@kfu.edu.sa}}$^{b}$

\vspace{5mm}

{$^{a}$\em Laboratory of Theoretical Physics,  
Faculty of Sciences, Choua\"ib Doukkali University},\\
{\em PO Box 20, 24000 El Jadida, Morocco}

{$^b$\em Physics Department, College of Science, King Faisal University,\\
 PO Box
380, Alahsa 31982, Saudi Arabia}





\vspace{3cm}

\begin{abstract}

We study two entropies 
of a system composed of two coupled harmonic oscillators
which is brought to a
canonical thermal equilibrium with a heat-bath at temperature
$T$. Using the purity function, we explicitly determine the R\'enyi and van Newmon
entropies in terms of different physical parameters. We will numerically analyze
these two entropies under suitable conditions and show their relevance.

\vspace{3cm}



\end{abstract}
\end{center}
\end{titlepage}


\section{Introduction}


The study of information carried by signals attracted several researchers because of its relevance to
telecommunications. 
Historically, the first theory on the subject going back to Shannon \cite{Shannon} who formulated 
 a mathematical tool based on the probability aspects of events and initiated a new field of
research called actually information theory. Indeed, Shannon showed that
the amount of
information  carried by a sequence of events $p_1, p_2 , \cdots$ can be described by the entropy
$S(p)= -K\sum_{i=1}^{N} p_i\ln p_i$ with $K$ a positive constant. It has to verify 
three conditions on: (i)
$S(P)$ should be continuous in $p_i$, (ii) $S(p)$ should be a monotonic increasing function of $N$
when all $p_i=\frac{1}{N}$ are equally probably, (iii) $S(p)$ should be additive.
 Later on, 
 the Shannon theory has been extended to many measures of information
or entropy.
One of them is due to R\'enyi \cite{Renyi2}, which he 
was able to extend the Shannon
entropy to a continuous family of entropies of the forms $S_q = \frac{\ln\Tr \rho^q}{1 - q}$, with a single
parameter $q>1$.  The entropies $S_q $ cover also that of the von Neumann $S_1$, which can be recovered 
by requiring 
the limit $q\lga 1$.

For a many-body quantum system composed of two subsystems $(A,B)$, 
the bipartite entanglement between subsystems
is described by
a state $\Psi$ of the Hilbert space ${\cal{H}}={\cal{H}}_A \otimes {\cal{H}}_B$. The corresponding 
the reduced density matrix  $\rho_A= \Tr_{B}\left(\rho_{AB}\right)$
 is obtained by tracing out the density matrix of the full system $\rho_{AB}=|\Psi\rangle\langle\Psi|$.
Noting that if $\rho_{AB}$   is a pure state then 
it suffices to use the von Neumann entropy in order to 
measure 
the amount
of the entanglement. 
However, the R\'enyi entropy has further importance
because it provides complete information about
the eigenvalue distribution of the reduced density matrix
$\rho_A$ and therefore completely characterizes the entanglement
in an overall pure, bipartite state \cite{21,22}. In fact,
the entanglement encodes the amount of non-classical information shared between 
complementary parts of an extended quantum state. For a pure state described by
density matrix, it can be quantified via the R\'enyi entanglement entropies.

In studying the entanglement in a quantum system, 
we have proposed a new approach \cite{Merdaci18} to explicitly determine the purity function
for the whole energy spectrum rather than the ground state as mostly used in the literature.
%
This was done
 by choosing the two coupled harmonic oscillators as system and using the path integral technique as tools
to deal with our issues. 
Among the obtained results, we have derived
a thermal wavefunction depending on temperature
%
that plied a crucial
role in discussing different properties of our system.
To prove the validity of our approach, we have showed that our results 
 reduce
 to 
the standard case describing the quantum system  in
the ground state at absolute zero temperature. This result has been obtained
in our previous work dealing with the 
entanglement in coupled harmonic
oscillators studied using a unitary
transformation
\cite{jellalstat}.

We  deal with other issues
related to the thermal wavefunction obtained in our work \cite{Merdaci18}. 
More precisely,
we study the two entropies corresponding to two coupled harmonic oscillators,
which is brought to a
canonical thermal equilibrium with a heat-bath at temperature
$T$.
%
%
%
Indeed, we use our purity function 
to explicitly
determine the R\'enyi  and von Neumann entropies as function of the temperature
parameter introduced through the path integral method. In fact, we show that
the von Neumann entropy can be derived as limiting case $q\lga 1$ of that of R\'enyi
of order $q$.
%
%
To highlight our results we present different density plots
of both entropies and show their basic properties.
These will be done by choosing different configurations of
the coupling parameter $\eta$, mixing angle $\theta$ and temperature $\beta=\frac{1}{k_BT}$.

The present paper is organized as follows. In section 2, 
we review our main results \cite{Merdaci18} needed
to deal with our task, which include the derivation of the reduced
density matrix and purity function for two coupled harmonic oscillators.
These will be used in section 3 to determine the R\'enyi entropies $S_q$ of all orders
$q$ as function of different physical parameters of our theory. We numerically focus
on $q=3$ to present some density plots showing the behavior of the entropy $S_3$.
In section 4, we consider the limit $q\lga 1$ to end up with the von Neumann
entropy $S_1$ as particular case. We give three tables chowing the particular forms of $S_1$
according to the nature of system at high and low temperature as well as some
plots will be presented.
We conclude our results in the final section.


\section{Thermal wavefunction}


To do our task we review the main results derived in our previous work \cite{Merdaci18} by
considering a system of two coupled harmonic
oscillators of masses $(m_1,m_2)$ parameterized by the planar
coordinates $(x_1,x_2)$. This system is described by the  Hamiltonian 
\cite{jellal}
\begin{equation}  \label{1}
\hat{H}=\frac{\hat{p}_{1}^{2}}{2m_{1}}+\frac{\hat{p}_{2}^{2}}{2m_{2}}+\frac {%
1}{2}C_{1}\hat{x}_{1}^{2}+\frac{1}{2}C_{2}\hat{x}_{2}^{2}+\frac{1}{2}C_{3}%
\hat{x}_{1}\hat{x}_{2}
\end{equation}
where $C_1, C_2$ and $C_3$ are constant parameters. It is clear that the
decoupled harmonic oscillators are recovered by requiring $C_3=0$. In the next, we will
adopt 
the path formalism to explicitly determine the thermal
wavefunction corresponding to the present system and later on derive the corresponding purity function.
In doing so, we proceed by
introducing the density matrix and particularly the reduced density matrix.

For imaginary time, 
the  propagator of a system is equivalent
to the density matrix for a particle that is in a heat bath.
Thus, the density matrix of the system
can be obtained directly from the propagator under an unitary transformation 
of angle
\beq\lb{theta}
\tan\theta=\frac{C_{3}}{\mu^{2}C_{2}-\frac{C_{1}}{\mu^{2}}},\qquad
\mu=\left( \frac{m_{1}}{m_{2}}\right) ^{\frac{1}{4}}.
\eeq
In constructing
the the path integral for the propagator corresponding to the  Hamiltonian \eqref{1},
according to \cite{Kosztin, rossi}
we consider the energy shift 
\beq
\hat{H}\longrightarrow \hat{H}-E_{0}\hat{\mathbb I}
\eeq
to ensure that the  wavefunction of the system converges to that
of the ground state at low temperature {($T\lga 0$)}. Now let us 
  introduce the evolution operator
\begin{equation}\lb{prop}
\mathbf{\hat{U}}(\beta)=\mathcal{T}_{D}\exp\left(  -\int_{0}^{\beta}\left(
\hat{H}-E_{0}\hat{\mathbb I}\right)  d\tau\right)  =e^{+\beta E_{0}}\mathcal{T}_{D}%
\exp\left(  -\int_{0}^{\beta}\hat{H}d\tau\right)
\end{equation}
with $\mathcal{T}_{D}$ being chronological Dyson
operator. 
{Because of the partition function does not determine any local thermodynamic quantities, 
then important local information resides in the thermal analog of the time evolution amplitude
\cite{Kleinert}}
\begin{equation}
\rho^{AB}(x_{1b},x_{2b},x_{1a},x_{2a};\beta)=\langle x_{1b},x_{2b}|\mathbf{%
\hat{U}}(\beta)| x_{1a},x_{2a}\rangle
\end{equation}
{
which are the matrix elements of the propagator \eqref{prop}},
where $A$ and $B$ are two subregions forming our system, 
with $\mid x_{1a},x_{2a}\rangle$ and $\mid x_{1b},x_{2b}\rangle$ are the initial 
and final states. 
In the forthcoming analysis, we consider the shorthand notation $%
\rho^{AB}(x_{1b},x_{2b},x_{1a},x_{2a};\beta)=\rho^{AB}\left( b,a;
\beta\right)$. 
Using the path integral method, 
we obtain
the density matrix elements 
\begin{eqnarray}  \label{rhoab1}
\rho^{AB}\left( b,a;\beta\right) & =& \frac{m\omega }{2\pi\hbar } e^{+\beta E_{0}}\left( \tfrac{1} {%
\sinh\left( \hbar\omega\beta e^{\eta}\right) \sinh\left( \hbar\omega\beta
e^{-\eta}\right) }\right) ^{\frac{1}{2}}\exp\left\{
-ax_{1b}^{2}-bx_{2b}^{2}-ax_{1a}^{2}-bx_{2a}^{2}\right\} \\
&& \times\exp\left\{
2cx_{1b}x_{2b}+2cx_{1a}x_{2a}+2dx_{1b}x_{1a}+2fx_{2b}x_{2a}-2gx_{1b}x_{2a}-2gx_{1a}x_{2b}\right\}
\notag
\end{eqnarray}
where different quantities are given by
\begin{eqnarray}
&&a=\mu^{2}\frac{m\omega }{2\hbar}\left[ {e^{\eta}\coth\left(
\hbar\omega\beta e^{\eta}\right) }\cos^{2} \tfrac{\theta}{2} +{e^{-\eta}\coth\left(
\hbar\omega\beta e^{-\eta}\right) }\sin^{2} \tfrac{\theta}{2} \right] \\
&& b=\frac{m\omega}{\mu^{2}2\hbar}\left[ {e^{\eta}\coth\left(
\hbar\omega\beta e^{\eta}\right) }\sin^{2} \tfrac{\theta}{2} +{e^{-\eta}\coth\left(
\hbar\omega\beta e^{-\eta}\right) }\cos^{2} \tfrac{\theta}{2} \right] \\
&& c=\frac{m\omega}{2\hbar}\left( {e^{\eta}\coth\left(
\hbar\omega\beta e^{\eta}\right) } -{e^{-\eta} \coth\left( \hbar\omega\beta
e^{-\eta}\right) } \right)
\cos \tfrac{\theta}{2} \sin \tfrac{\theta}{2} \\
&& d=\frac{\mu^{2}m\omega}{2\hbar}\left[ \tfrac{e^{\eta}}{\sinh\left(
\hbar\omega\beta e^{\eta}\right) }\cos^{2} \tfrac{\theta}{2} +\tfrac{%
e^{-\eta}}{\sinh\left( \hbar\omega\beta e^{-\eta }\right) }\sin^{2} \tfrac{%
\theta}{2} \right] \\
&& f=\frac{m\omega}{\mu^{2}2\hbar}\left[ \tfrac{e^{\eta}}{\sinh\left(
\hbar\omega\beta e^{\eta}\right) }\sin^{2} \tfrac{\theta}{2} +\tfrac{%
e^{-\eta}}{\sinh\left( \hbar\omega\beta e^{-\eta }\right) }\cos^{2} \tfrac{%
\theta}{2} \right] \\
&& g=\frac{m\omega}{2\hbar}\left( \tfrac{e^{\eta}}{\sinh\left( \hbar
\omega\beta e^{\eta}\right) }-\tfrac{e^{-\eta}}{\sinh\left( \hbar\omega\beta
e^{-\eta}\right) }\right) \cos \tfrac{\theta}{2} \sin \tfrac{\theta}{2}
\end{eqnarray}
and we have set the coupling parameter 
\begin{equation}
\lb{eta}
e^{\pm2\eta}=\frac{\frac{C_{1}}{\mu^{2}}+\mu^{2}C_{2}\mp\sqrt{\left( \frac{%
C_{1}}{\mu^{2}}-\mu^{2}C_{2}\right) ^{2}+C_{3}^{2}}}{2k}
\end{equation}
as well as the frequency  $\omega=\sqrt{\frac{k}{m}}$ with the mass $m=\sqrt{m_{1}m_{2}}$ and the 
coupling strength $k=\sqrt{C_{1}C_{2}-\frac{C_{3}^{2}}{4}}$.


To derive the thermal wavefunction associated to the Hamiltonian \eqref{1}, we 
make use 
the variable substitution $\left( x_{1a},x_{2a}\right) =\left(
x_{1b},x_{2b}\right) =\left( x_{1},x_{2}\right)$ into \eqref{rhoab1} and take only the diagonal elements
of  the density matrix. These allow to get the probability density
\begin{equation}  \label{eqq20}
P_\beta(x_{1},x_{2})=\mbox{diag} \left(\rho^{AB}(b,a;%
\beta) \right)
\end{equation}
and explicitly we have 
\begin{equation}\lb{gstate}
P_\beta(x_{1},x_{2})= \frac{m\omega e^{+\beta E_{0}}}{2\pi\hbar \sqrt{\sinh\left(
\hbar\omega\beta e^{\eta}\right) \sinh\left( \hbar\omega\beta
e^{-\eta}\right) }} e^{-\tilde{a}(\beta)x_{1}^{2}-\tilde{b}(\beta)x_{2}^{2}+2%
\tilde{c}(\beta)x_{1}x_{2}}
\end{equation}
where  the shorthand notations are used
\begin{align}
\tilde{a}(\beta) & =2(a-d)=\mu^{2} \tfrac{m\omega}{\hbar}\left[
e^{\eta}\tanh\left( \frac{\hbar\omega}{2}\beta e^{\eta}\right) \cos ^{2}
\tfrac{\theta}{2} +e^{-\eta}\tanh\left( \frac{\hbar\omega} {2}\beta
e^{-\eta} \right) \sin^{2} \tfrac{\theta}{2} \right] \\
\tilde{b}(\beta) & =2(b-f)=\tfrac{m\omega}{\mu^{2}\hbar}\left[ e^{\eta}\tanh%
\left( \frac{\hbar\omega}{2}\beta e^{\eta}\right) \sin ^{2} \tfrac{\theta}{2}
+e^{-\eta}\tanh\left( \frac{\hbar\omega}{2}\beta e^{-\eta}\right) \cos^{2}
\tfrac{\theta}{2} \right] \\
\tilde{c} (\beta)& =2(c-g)=\tfrac{m\omega}{\hbar}\left[ e^{\eta}\tanh\left(
\frac{\hbar\omega}{2}\beta e^{\eta}\right) -e^{-\eta}\tanh\left( \frac{%
\hbar\omega}{2}\beta e^{-\eta}\right) \right] \cos \tfrac{\theta}{2} \sin
\tfrac{\theta}{2}.
\end{align}

Generally for any temperature parameter $\beta$,  the wavefunction describing  our system can be determined 
by integrating over the initial variables as has been done in \cite{Kosztin}. Thus, in our case we have to
write the solution of the imaginary time Schr\"odinger equation as
\beq\lb{gfun}
\psi(x_{1},x_{2};\beta) = \int \rho^{AB} \left(b,a;\beta-\frac{\varepsilon}{2}\right) 
\psi\left(x_{1a},x_{2a};\frac{\varepsilon}{2}\right) \ dx_{1a} dx_{2a}
\eeq
where the density matrix of the system verifies the condition
\beq
\underset{\beta\longrightarrow\frac{\varepsilon}{2}}{\lim } \ \rho^{AB}\left(b,a;\beta-\frac{\varepsilon}{2}\right) =
\del(x_1-x_{1a}) \del(x_2-x_{2a}).
\eeq
and the non-normalized  initial  wavefunction is 
\beq\lb{000}
\psi\left(x_{1a},x_{2a};\frac{\varepsilon}{2}\right)=  
\tfrac{\sqrt
{\tfrac{m\omega}{4\pi\hbar}}}{\sqrt{\cosh\left(  {\hbar\omega e^{\eta}%
}\frac{\varepsilon}{2}\right)  \cosh\left(  {\hbar\omega e^{-\eta}}%
\frac{\varepsilon}{2}\right)  }}
e^{-\frac{1}{2}\tilde{a}(\varepsilon)x_{1}^{2}-\frac{1}{2}\tilde{b}(\varepsilon)x_{2}^{2}
+\tilde{c}(\varepsilon)x_{1}x_{2}}
\eeq
where $\varepsilon$ is a small value of the high temperature, 
which is introduced
to insure 
{the convergence of the probability density  of the initial state.}
Now replacing \eqref{000} and integrating \eqref{gfun} to show that  the thermal wavefunction
takes the form
\bqr\lb{333}
\psi(x_{1},x_{2};\beta)  = \tfrac{\sqrt
{\tfrac{m\omega}{4\pi\hbar}}}{\sqrt{\cosh\left(  {\hbar\omega e^{\eta}%
}\beta\right)  \cosh\left(  {\hbar\omega e^{-\eta}}%
\beta\right)  }}
\ e^{+\beta \hbar\omega\cosh\eta}
e^{-\tilde{\alpha} x_{1}^{2}-\tilde{\beta} x_{2}^{2}+2\tilde{\gamma} x_{1}x_{2}}
\eqr
where we have set the quantities 
\begin{align}
\tilde{\alpha} & =\mu^{2} \tfrac{m\omega}{2\hbar}\left[
e^{\eta}\tanh\left( {\hbar\omega} e^{\eta} \beta\right) \cos ^{2}
\tfrac{\theta}{2} +e^{-\eta}\tanh\left( {\hbar\omega} 
e^{-\eta} \beta \right) \sin^{2} \tfrac{\theta}{2} \right]\lb{alpha} \\
\tilde{\beta} & =\tfrac{m\omega}{2\mu^{2}\hbar}\left[ e^{\eta}\tanh%
\left( {\hbar\omega} e^{\eta} \beta \right) \sin ^{2} \tfrac{\theta}{2}
+e^{-\eta}\tanh\left( {\hbar\omega}  e^{-\eta} \beta \right) \cos^{2}
\tfrac{\theta}{2} \right]\lb{beta} \\
\tilde{\gamma} & =\tfrac{m\omega}{2\hbar}\left[ e^{\eta}\tanh\left(
{\hbar\omega} e^{\eta} \beta \right) -e^{-\eta}\tanh\left( {%
\hbar\omega}e^{-\eta} \beta \right) \right] \cos \tfrac{\theta}{2} \sin
\tfrac{\theta}{2}.\label{gamma}
\end{align}
It is interesting to underline that firstly $\psi(x_{1},x_{2};\beta)$ is temperature dependent and satisfies
the imaginary time Schr\"odinger equation
\beq\lb{schro}
\left(  \hat{H}-E_{0}\hat{\mathbb I}\right) \psi(x_{1},x_{2};\beta) + \frac{\pa}{\pa\beta}\psi(x_{1},x_{2};\beta)=0
\eeq
where 
the substitution $t\lga -i\hbar \beta$ is taken into account. This clearly shows the reason behind taking
the energy shift in the Hamiltonian system. Secondly $\psi(x_{1},x_{2};\beta)$ is the wavefunction
corresponding the whole energy spectrum. This issue and related matters were discussed in our previous work
\cite{Merdaci18}
dealing with the entanglement of our system.


Now we have obtained all ingredients to do our tasks. Indeed, 
using the
standard definition based on the thermal wavefunction 
\begin{equation}
\rho_{\mathsf{red}}^{A}(x_{1},x_{1}^{\prime};\beta) =
\frac{\int\psi(x_{1},x_{2};\beta)\psi^{\ast}(x_{1}^{\prime},x_{2}%
;\beta)dx_{2}}{\int\psi(x_{1},x_{2};\beta)\psi^{\ast}(x_{1},x_{2}%
;\beta)dx_{1}dx_{2}}
\label{313}
\end{equation}
to end up with the reduced density matrix 
\begin{eqnarray}
\rho_{\mathsf{red}}^{A}(x_{1},x_{1}^{\prime};\beta) 
=  \sqrt{2\frac{\tilde{\alpha}\tilde{\beta}-\tilde{\gamma}^{2}}
{\pi\tilde {\beta}}}
\exp\left( {-\frac{2%
\tilde{\alpha}\tilde{\beta}-\tilde{\gamma}^{2}}
{2\tilde{\beta}}x_{1}^{2}-%
\frac{2\tilde{\alpha}\tilde{\beta}-\tilde{\gamma }^{2}}{2\tilde{\beta}}%
x_{1}^{\prime2}+\frac{\tilde{\gamma}^{2}}{\tilde{\beta}}x_{1}x_{1}^{\prime}}%
\right). \label{eqred}
\end{eqnarray}
We can do the same job to obtain 
a similar reduced density matrix  $\rho_{\mathsf{red}}^{B}$ 
of the subregion $B$ that can be
determined
%
by integrating \eqref{eqq20} over the variable $x_1$. These tell us that
for both subregions $A$ and $B$  
the  purity function  is the same $P^{A}=P^{B}=P$. 
It is defined as 
trace over square of the reduced density matrix \eqref{eqred} 
\begin{eqnarray}
P =\Tr_{A}\left( \rho_{\mathsf{red}}^{A}(x_{1},x_{1}^{\prime};\beta)\right) ^{2}
\end{eqnarray}
which can be calculated to get
\begin{eqnarray}
P=
\sqrt{\tfrac{\tanh\left( { \hbar\omega
} \beta e^{\eta} \right) 
\tanh\left( { \hbar\omega} \beta e^{-\eta}\right) } 
{ \left(e^{\eta}\tanh\left( { \hbar\omega}
\beta e^{\eta}\right) \sin^{2} \frac{\theta}{2} +e^{-\eta }\tanh\left( {
\hbar\omega} \beta e^{-\eta}\right) \cos^{2} \frac{\theta}{2} \right) \left(
e^{\eta}\tanh\left( { \hbar\omega} \beta e^{\eta}\right) \cos^{2}
\frac{\theta}{2} +e^{-\eta}\tanh\left( { \hbar\omega} \beta
e^{-\eta}\right) \sin^{2} \frac{\theta}{2} \right)}} \label{35}
\end{eqnarray}
as function of the coupling parameter 
$\eta$, mixing $%
\theta$ and temperature $\beta$. Note that the purity function $P$ is  the product of two quantities and they are
differentiating by the $\eta$ sign of the numerator and the geometric functions in the denominator.
Moreover, we notice that the derivation of such $P$  is actually based on exact calculation without use of 
approximation and it is corresponding to the whole energy spectrum.

Right now we have settled the need materials to do our task and 
next we see how to use them in order to determine some interesting quantities those
measure the amount of information for a given system. 
More precisely,
because of the purity function is linked to some entropies, then we will show that two entropies
can be derived from our results. These will be analyzed according to choice of different configurations
of the coupling parameter, mixing angle and temperature.

\section{R\'enyi entropy}

The definition of entropy does not in any way require the notion of an observer, but 
requires one has to specify the subspace of the system under consideration in order 
to get the density matrix.
An observer may measure different entropies depending on which aspects of the system is considered.
Concretely, for a system of two entangled particles one will measure 
different entropies for each of the particles independently than the full entangled state.
%
In general, the lack of information or the mixedness about the preparation of
a given state, can be quantified by using generalized entropic measures, such as the
R\'{e}nyi entropy \cite{Renyi}
\begin{equation}  \label{renye}
S_q = \frac{\ln\Tr \rho^q}{1 - q}
\end{equation}
and the Bastiaans-Tsallis entropy \cite{Bastiaans, Tsallis}
\begin{equation}
S_q^{BT} = \frac{1-\Tr \rho^q}{q - 1}
\end{equation}
where the parameter verifies the condition $q>1$. 
It is clearly seen that they 
 have 
 two interesting limiting cases.  Indeed, 
for $q = 2$ and from $S_q^{BT}$ we recover the well-known 
 linear entropy 
\beq
S_2^{BT}=S_L = 1 - \Tr \rho^2
\eeq
which has
range between zero associated to a completely pure state and $(1 - 1/d)$ associated 
to a completely mixed state, with $d$ is the dimension of the density matrix $\rho$.
Note that, the linear entropy is trivially related to the purity function $P$ of a state 
via 
$S_L= 1-P.$
For the limit $q\lga 1$, we end up with the
von Neumann entropy 
\begin{equation}
S_V= \lim_{q\longrightarrow 1^+} S_q^{BT} = \lim_{q\longrightarrow 1^+} S_q = - \Tr %
(\rho \ln \rho) 
\end{equation}
which  
is additive on tensor product states and provides a further convenient
measure of mixedness of the quantum state.

Having obtained the purity function $P$, 
let us  show how to drive   the  R\'{e}nyi and
 von Neumann entropies for two coupled harmonic oscillators
In doing so, we need first to write 
$\rho_{\mathsf{red}}^{A}(x_{1},x_{1}^{\prime};\beta)$ %
\eqref{eqred} in the Gaussian form as
\begin{equation}\lb{rhog}
\rho_{\mathsf{red}}^{A}(x_{1},x_{1}^{\prime};\beta)
=A \ e^{-ax_{1}^{2}-ax_{1}^{\prime2}+bx_{1}
x_{1}^{\prime}}
\end{equation}
where we have set the quantities
\begin{equation}
A= \sqrt{2\frac{\tilde{\alpha}\tilde{\beta}-\tilde{\gamma}^{2}}
{\pi\tilde {\beta}}}, \qquad 
a =\frac{2\tilde\alpha\tilde\beta-\tilde\gamma^{2}}{2\tilde\beta}, \qquad b=\frac{%
\tilde\gamma^{2}}{\tilde{\beta}}.
\end{equation}
and $\tilde{\alpha},\tilde{\beta}, \tilde\gamma$ are given in (\ref{alpha}-\ref{gamma}).
Note that,
this Gaussian form was studied in \cite{Adesso2,Pipek} by  dealing with
the measures of spatial entanglement in a two-electron model atom. Now tracing  \eqref{rhog} 
to end up with
\begin{equation}
\Tr\left(\rho^A_{\sf red}\right)^{q} =\frac{\left( 2P\right) ^{q}}{\left( 1+P\right) ^{q}-\left(
1-P\right) ^{q}}
\end{equation}
in terms of the purity function $P$ \eqref{35}. Replacing in \eqref{renye}
to get the R\'{e}nyi entropy corresponding to our system
\begin{equation}  \label{renyer}
S_{q}=\frac{q}{1-q}\ln\left( 1-\frac{1-P}{1+P}\right) -\frac{1}{1-q}%
\ln\left( 1-\left( \frac{1-P}{1+P}\right) ^{q}\right)
\end{equation}
which is similar to that obtained in \cite{Adesso} by studying the 
extremal entanglement and mixedness in continuous variable systems.
For
$q=2$, then  the R\'{e}nyi entropy \eqref{renyer} reduces to that of order 2
\begin{equation}
S_{2}= -\ln P
\end{equation}
and explicitly it is
\bqr
S_{2} &=& \frac{1}{2}
\ln
\left(e^{\eta}\tanh\left( { \hbar\omega}
\beta e^{\eta}\right) \sin^{2} \frac{\theta}{2} +e^{-\eta }\tanh\left( {
\hbar\omega} \beta e^{-\eta}\right) \cos^{2} \frac{\theta}{2} \right) \nn\\
&& + \frac{1}{2}
\ln \left(
e^{\eta}\tanh\left( { \hbar\omega} \beta e^{\eta}\right) \cos^{2}
\frac{\theta}{2} +e^{-\eta}\tanh\left( { \hbar\omega} \beta
e^{-\eta}\right) \sin^{2} \frac{\theta}{2} \right)\nn\\
&& - \frac{1}{2} \ln
\tanh\left( { \hbar\omega
} \beta e^{\eta} \right) - \frac{1}{2} \ln 
\tanh\left( { \hbar\omega} \beta e^{-\eta}\right).
\eqr

\begin{figure}[H]
\centering  \includegraphics[width=8cm,  height=5.1cm]{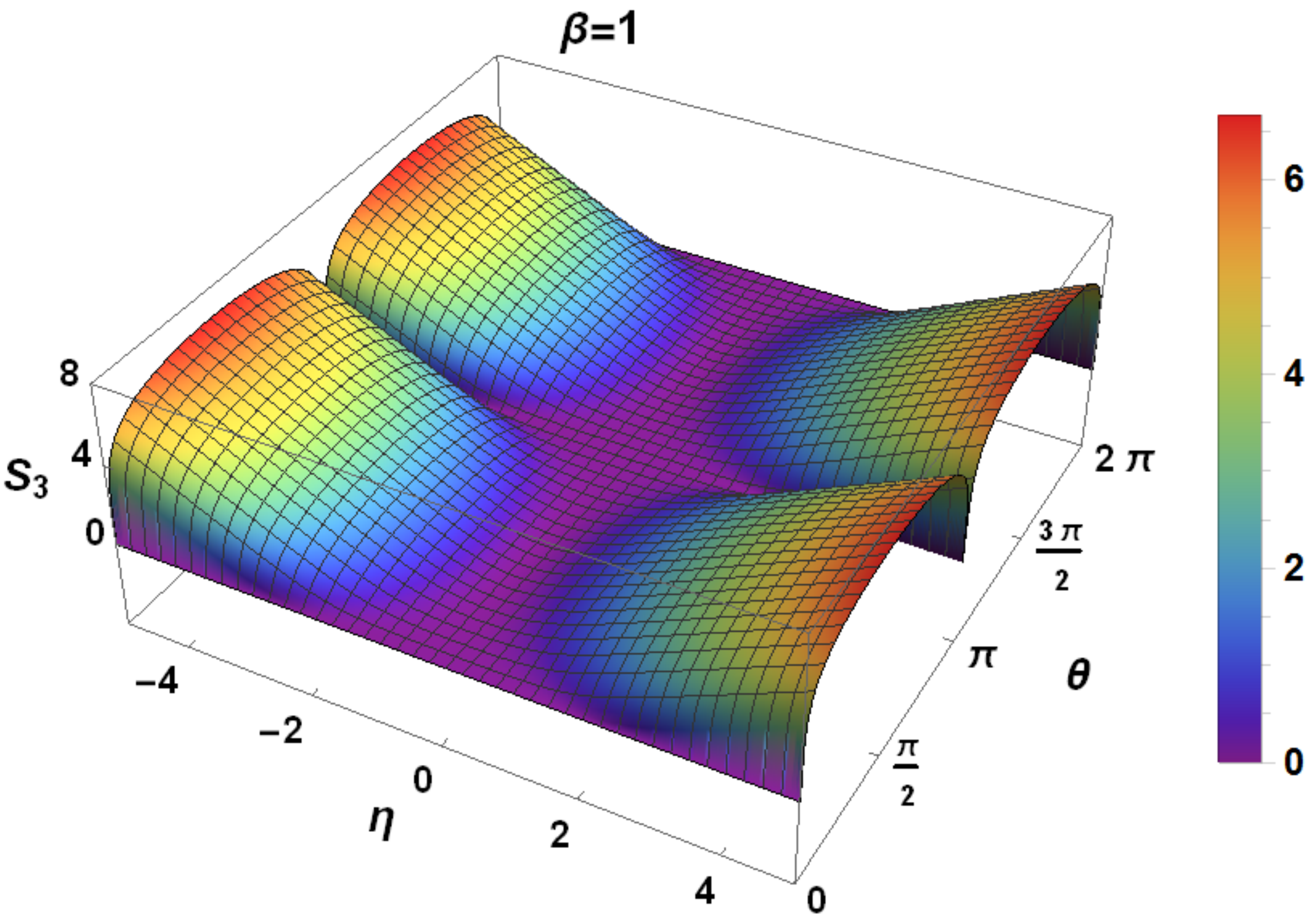}  %
\includegraphics[width=8cm,  height=5.1cm]{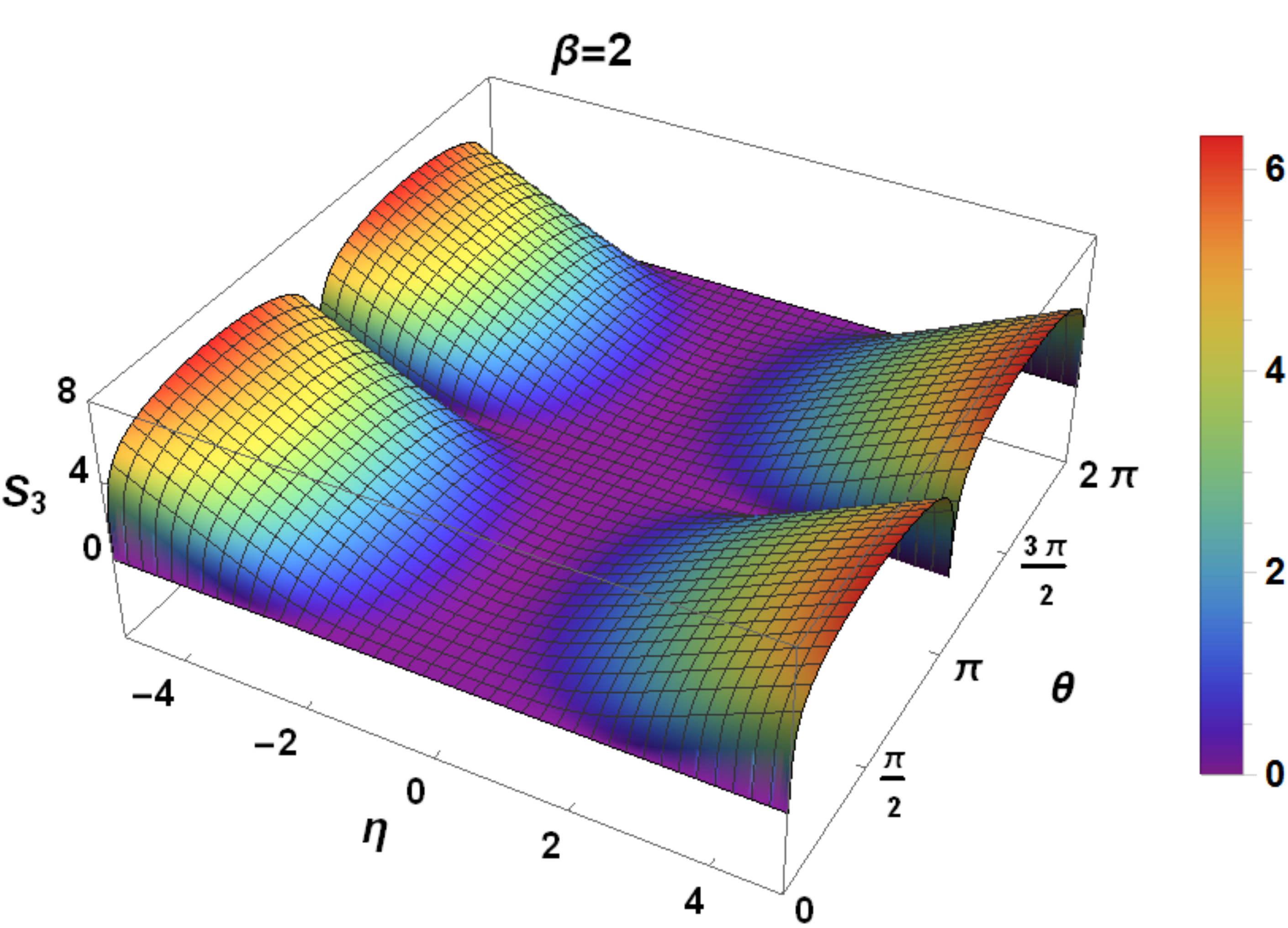}  %
\includegraphics[width=8cm,  height=5.1cm]{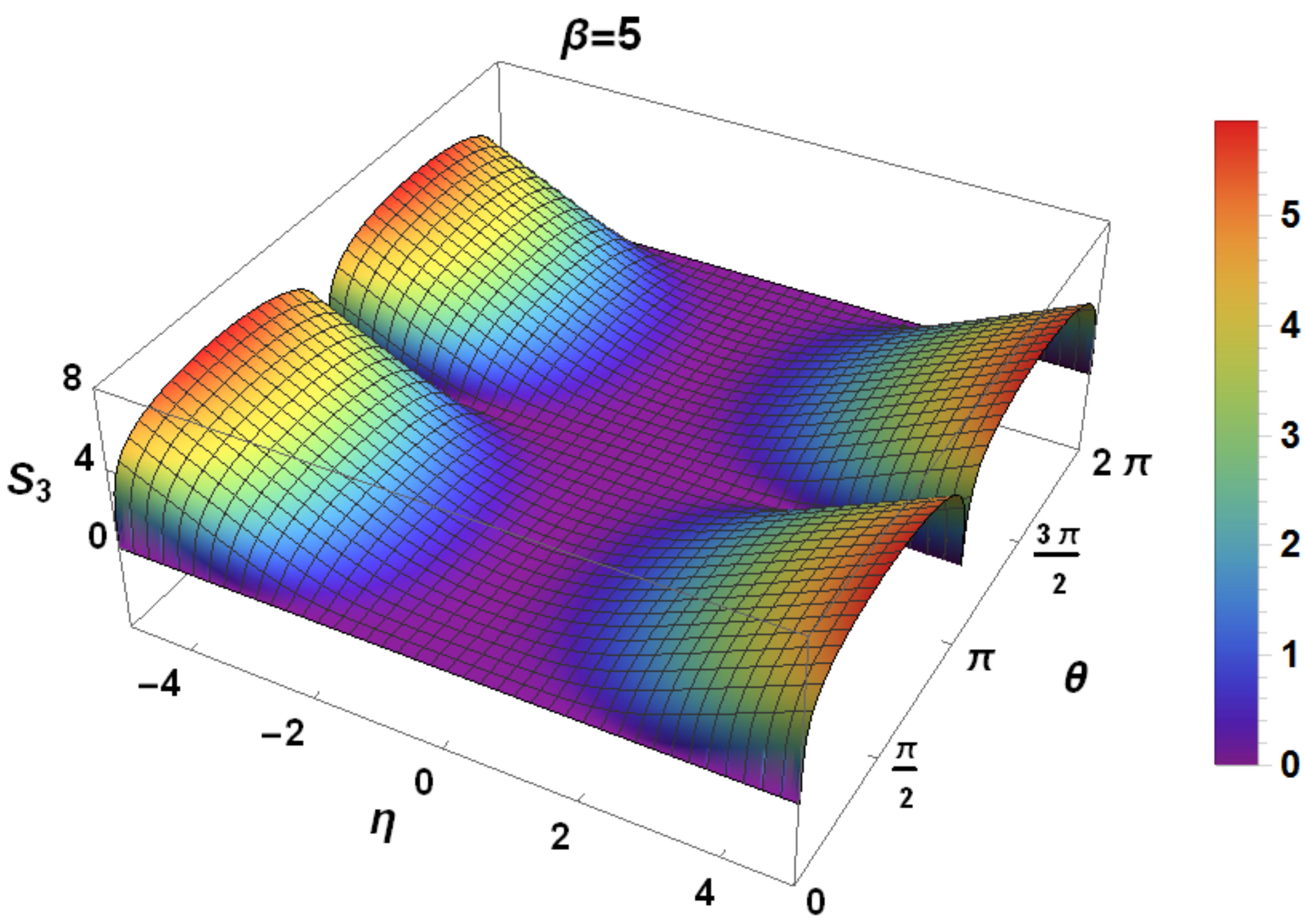}  %
\includegraphics[width=8cm,  height=5.1cm]{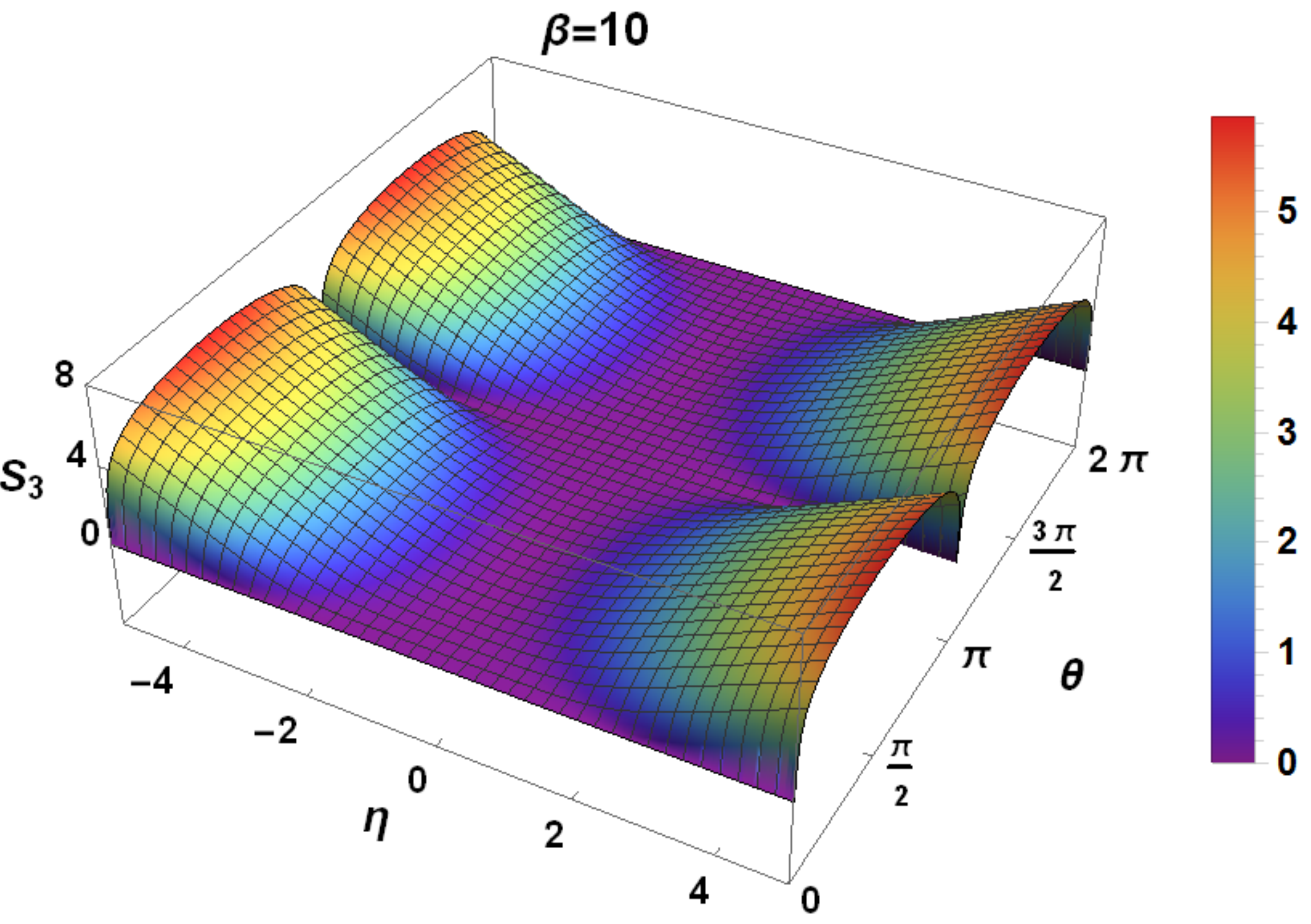}
\caption{\textsf{R\'{e}nyi entropy $S_3$ versus the coupling parameter $\protect\eta$ and the
mixing angle $\protect\theta$ for  fixed values of the temperature $\protect\beta=1,2,5,10$. }}
\label{R52-beta}
\end{figure}

At this stage, we can numerically analyze the R\'{e}nyi entropies and underline
their behaviors by choosing some configurations of the physical
parameters. For numerical difficulties, we restrict ourselves to
the entropy $S_3$, which can be obtained simply by fixing $q=3$
in \eqref{renyer}
\beq
S_3= \frac{1}{2}\ln \frac{3+P^2}{4P^2} 
\eeq
Figure \ref{R52-beta} presents 
the R\'{e}nyi entropy $S_3$ versus the coupling parameter $\protect\eta$ and the
mixing angle $\protect\theta$ for  fixed values of the temperature $\protect\beta=1,2,5,10$.
We observe that the entropy $S_3$ is periodic with respect to the mixing angle $\theta =\pi$
and increases from minimal to maximal values. Also $S_3$ shows a symmetric behavior with respect
to $\eta=0$ and it is null for a given interval of $\eta$ independently to the values
taken by $\theta$. 
This behavior changes as long as the temperature is
decreased from $\beta=1$ to $\beta=10$. This tell us how the temperature can be used to
control the behavior of our system and therefore it offers another way
to handle its correlations.

\begin{figure}[H]
\centering  \includegraphics[width=8cm,  height=5.1cm]{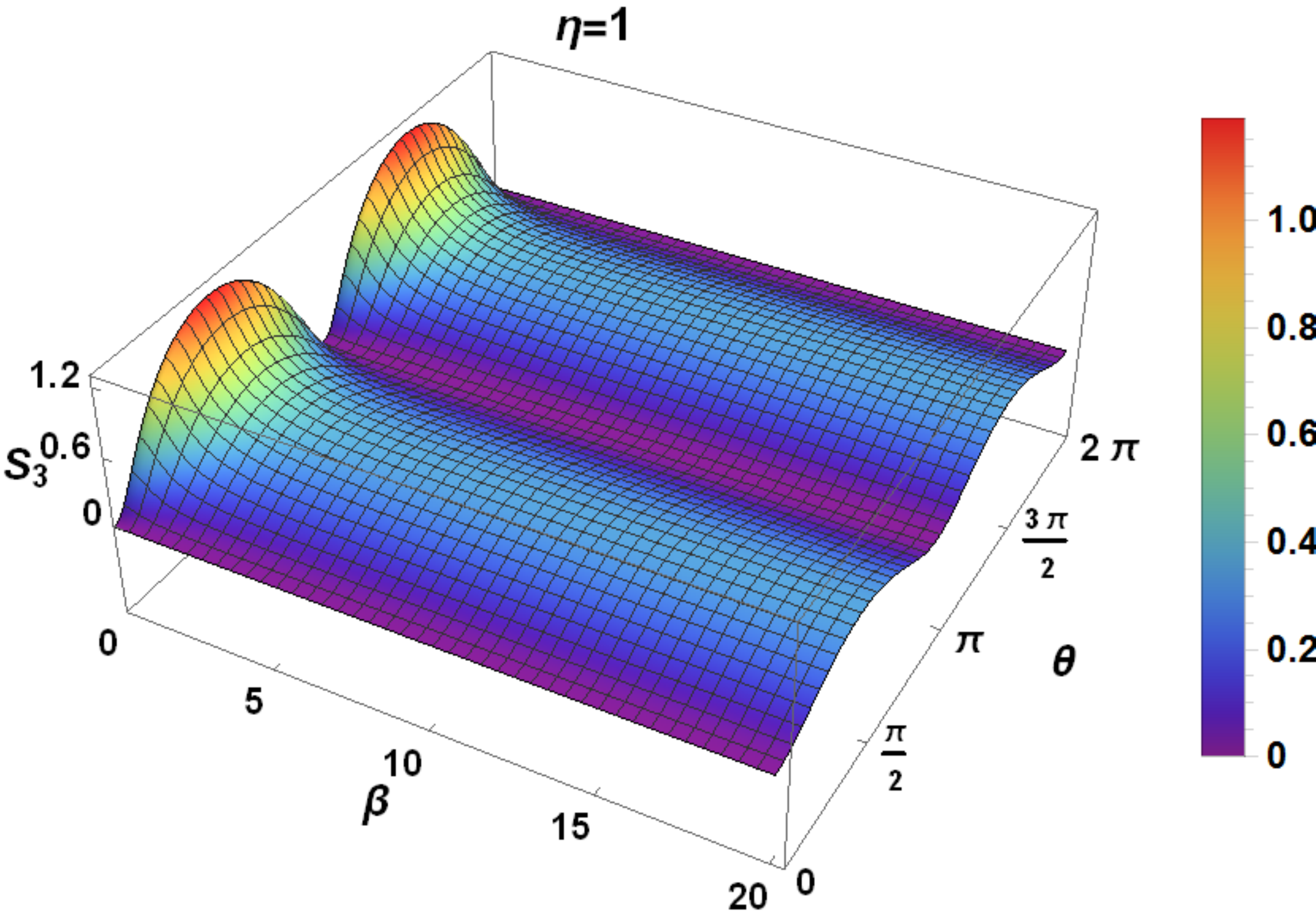}  %
\includegraphics[width=8cm,  height=5.1cm]{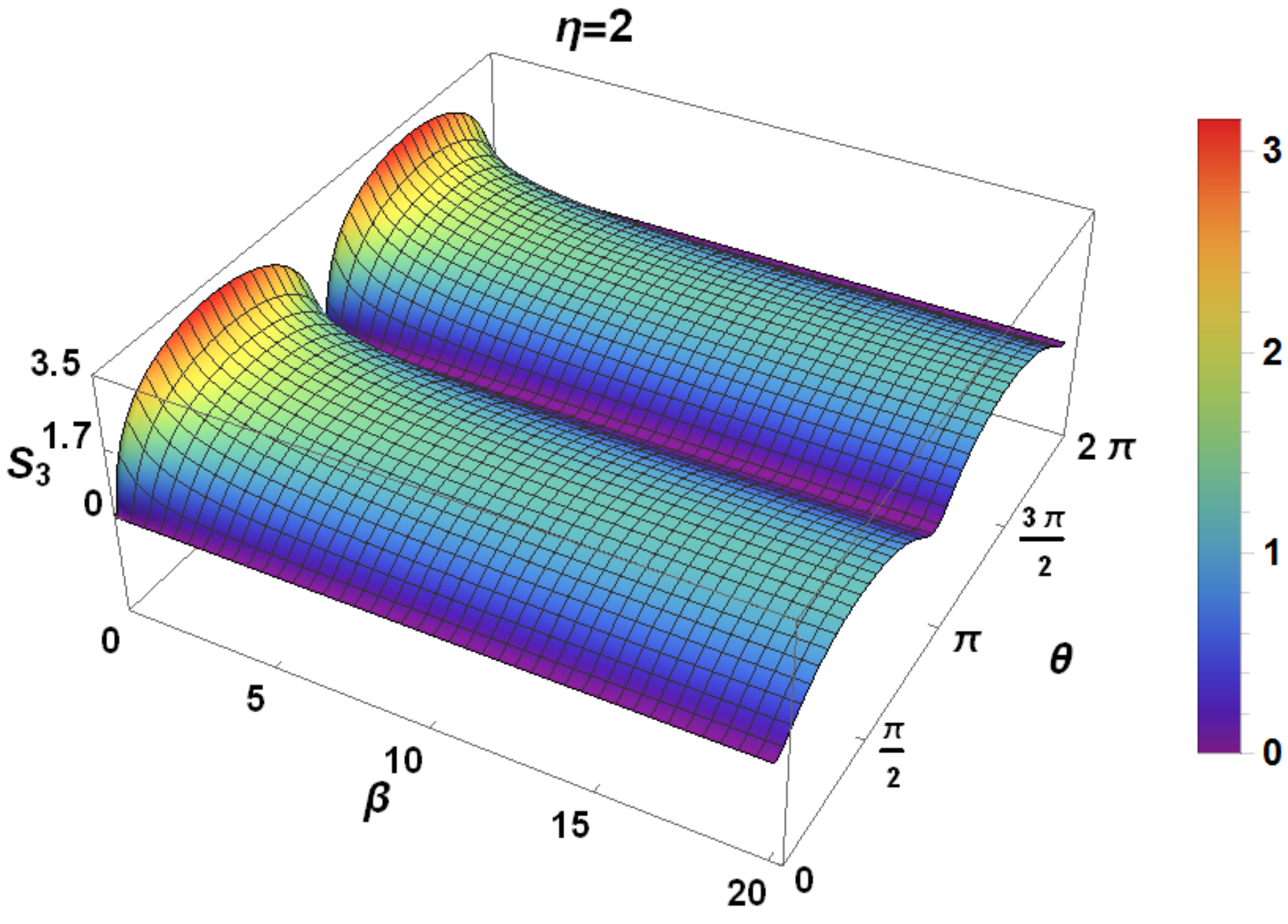}  %
\includegraphics[width=8cm,  height=5.1cm]{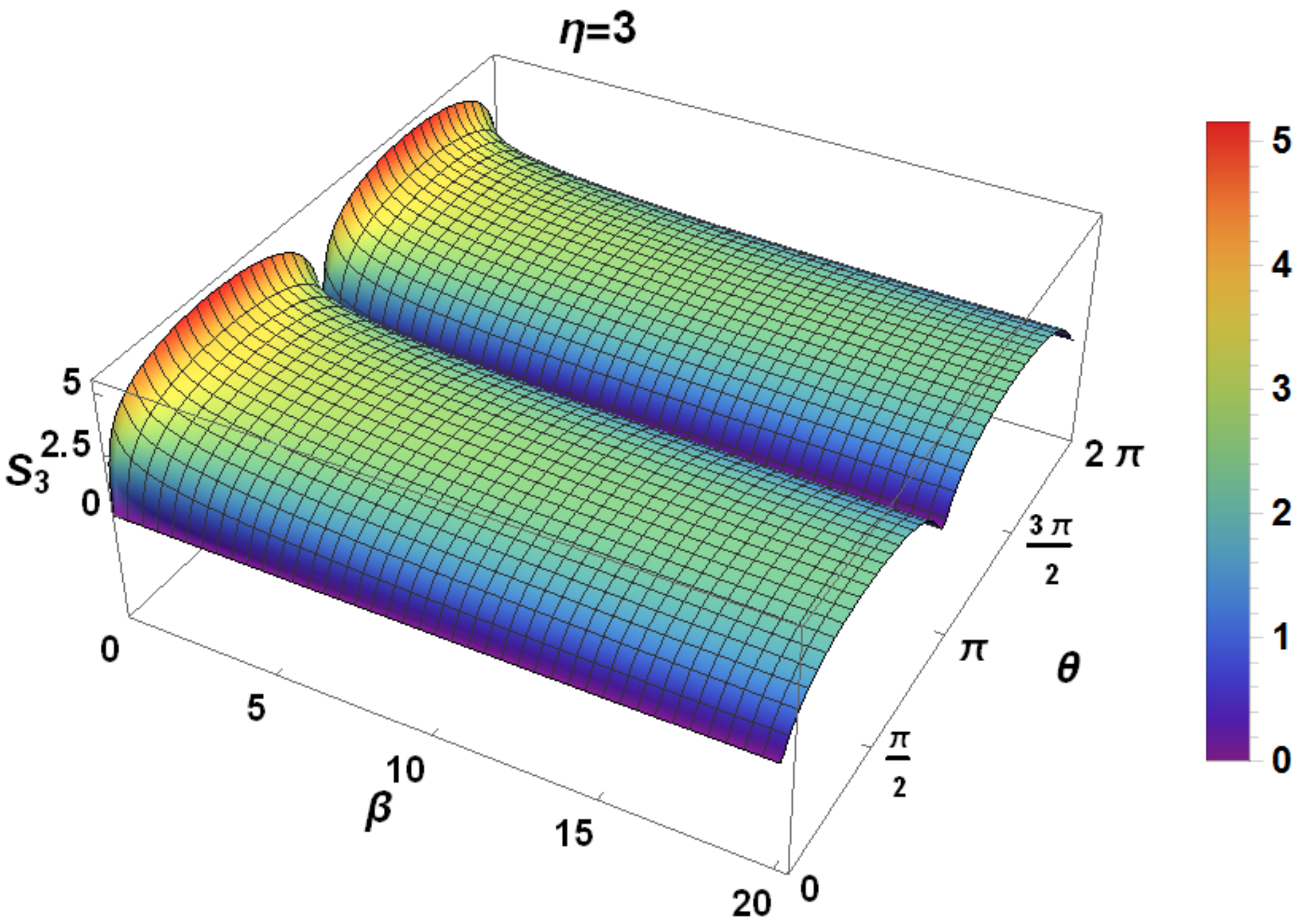}  %
\includegraphics[width=8cm,  height=5.1cm]{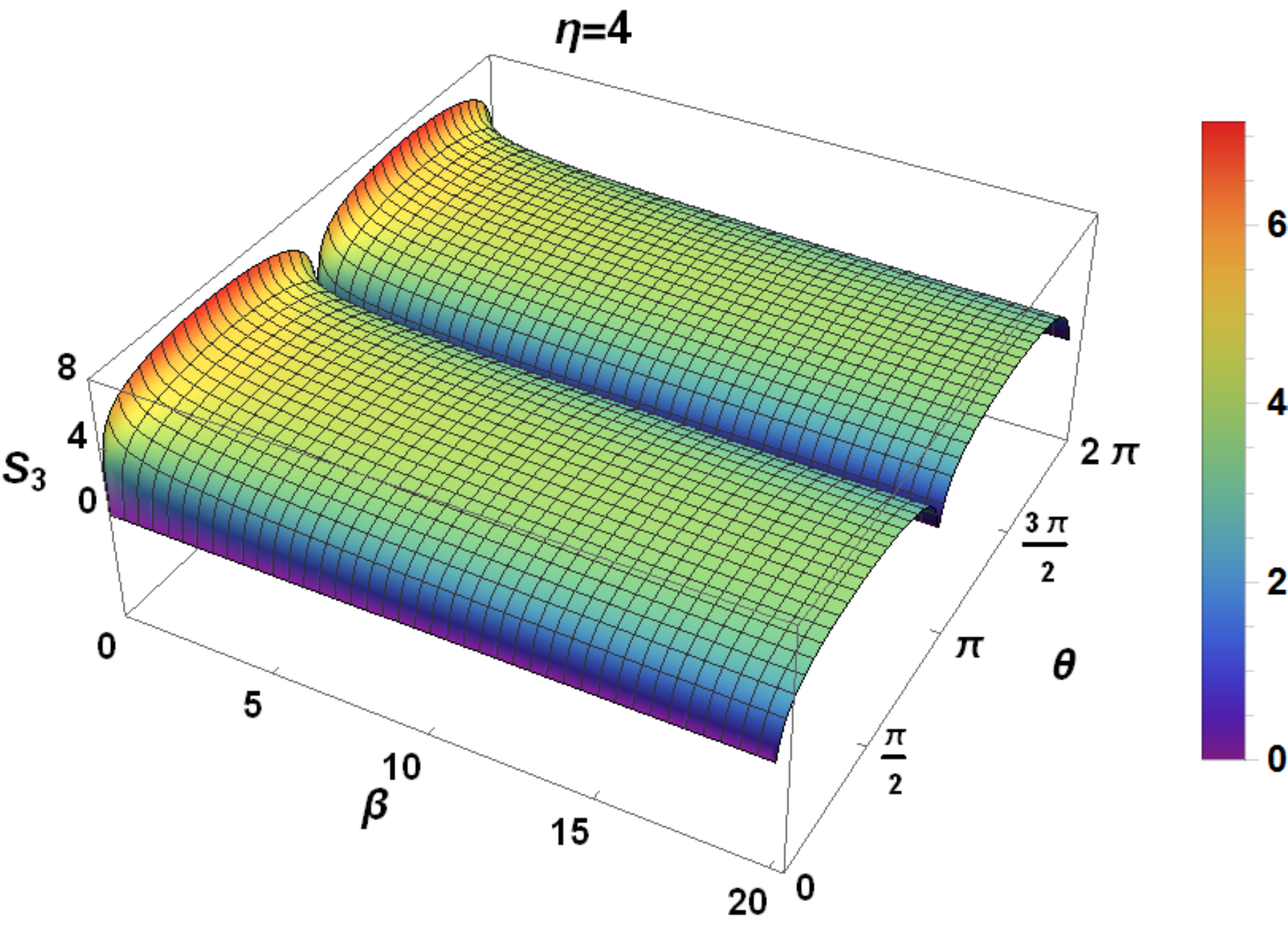}
\caption{\textsf{R\'{e}nyi entropy $S_3$ versus the temperature $\beta$ and the
mixing angle $\protect\theta$ for  fixed values of the coupling parameter $\eta=1,2,3,4$. }}
\label{R52-eta}
\end{figure}

\begin{figure}[H]
\centering  \includegraphics[width=8cm,  height=5.1cm]{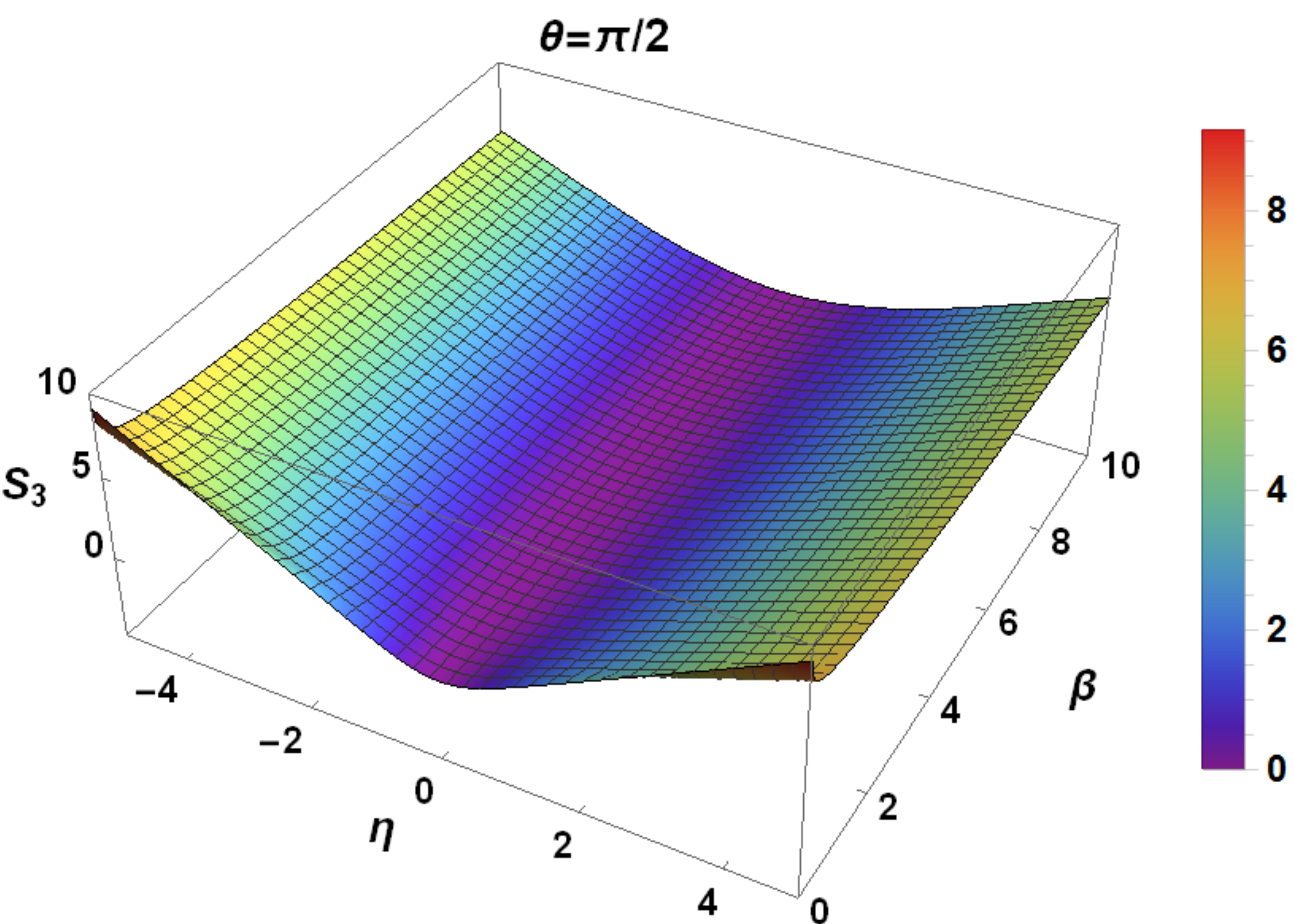}  %
\includegraphics[width=8cm,  height=5.1cm]{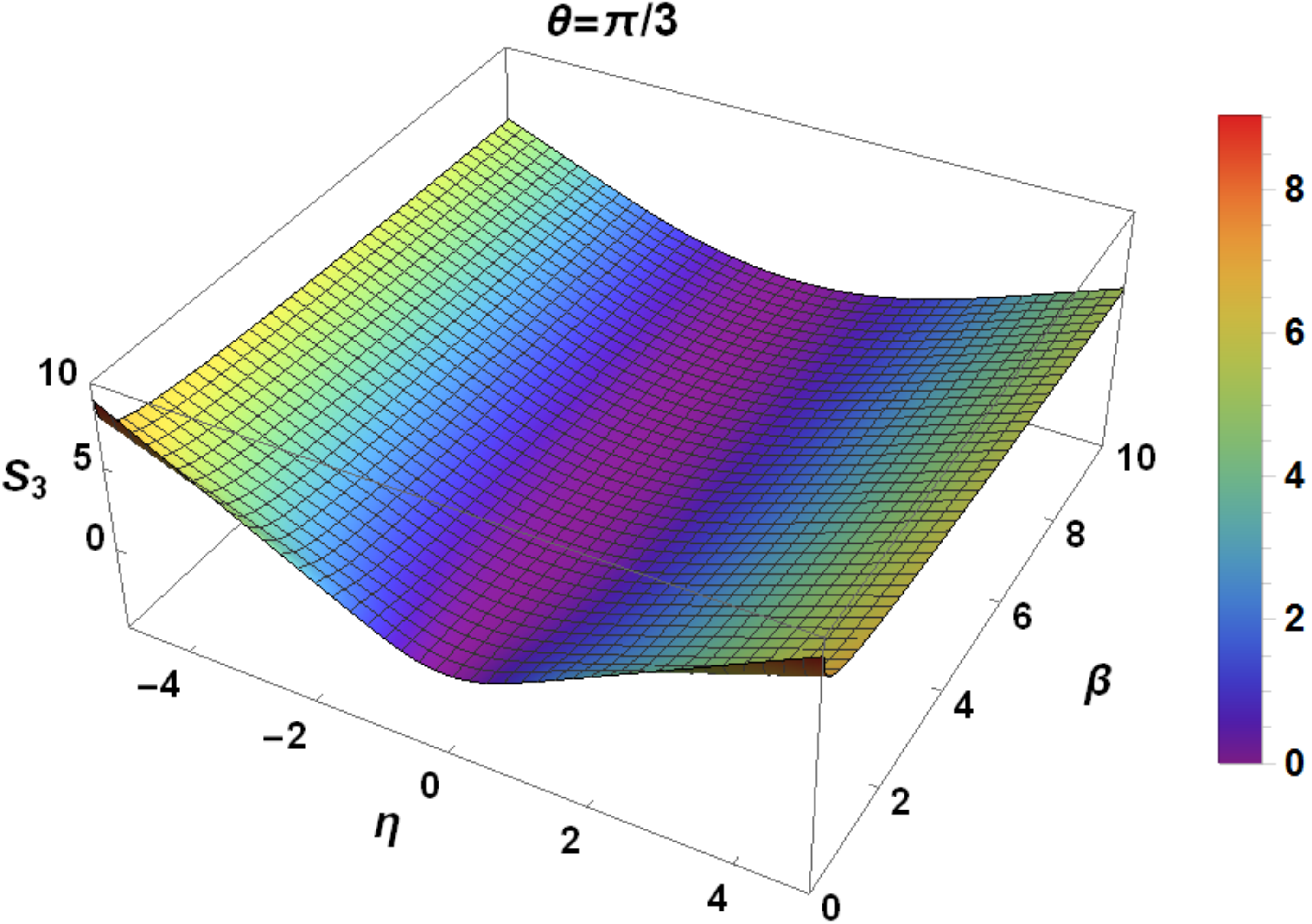}  %
\includegraphics[width=8cm,  height=5.1cm]{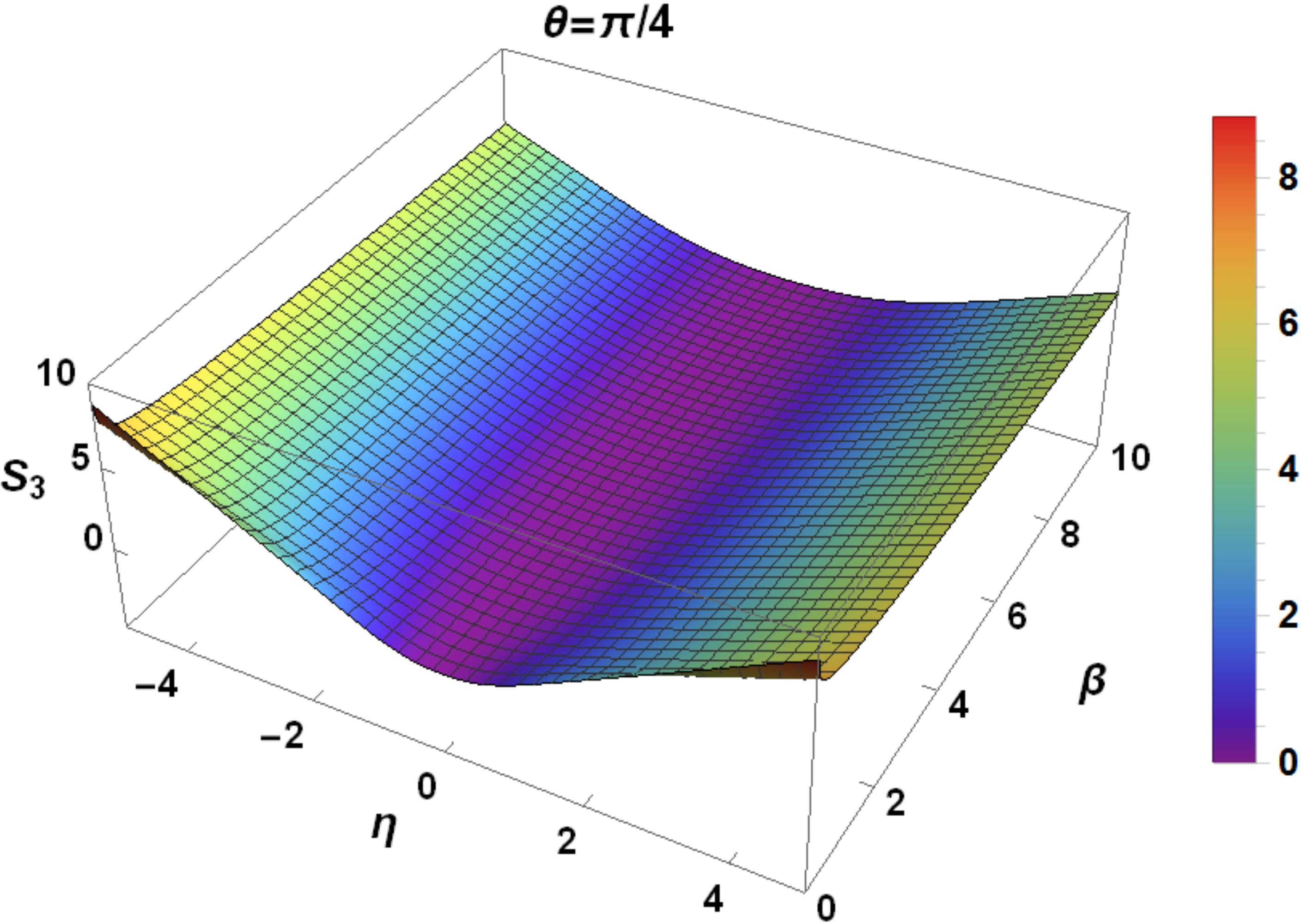}  %
\includegraphics[width=8cm,  height=5.1cm]{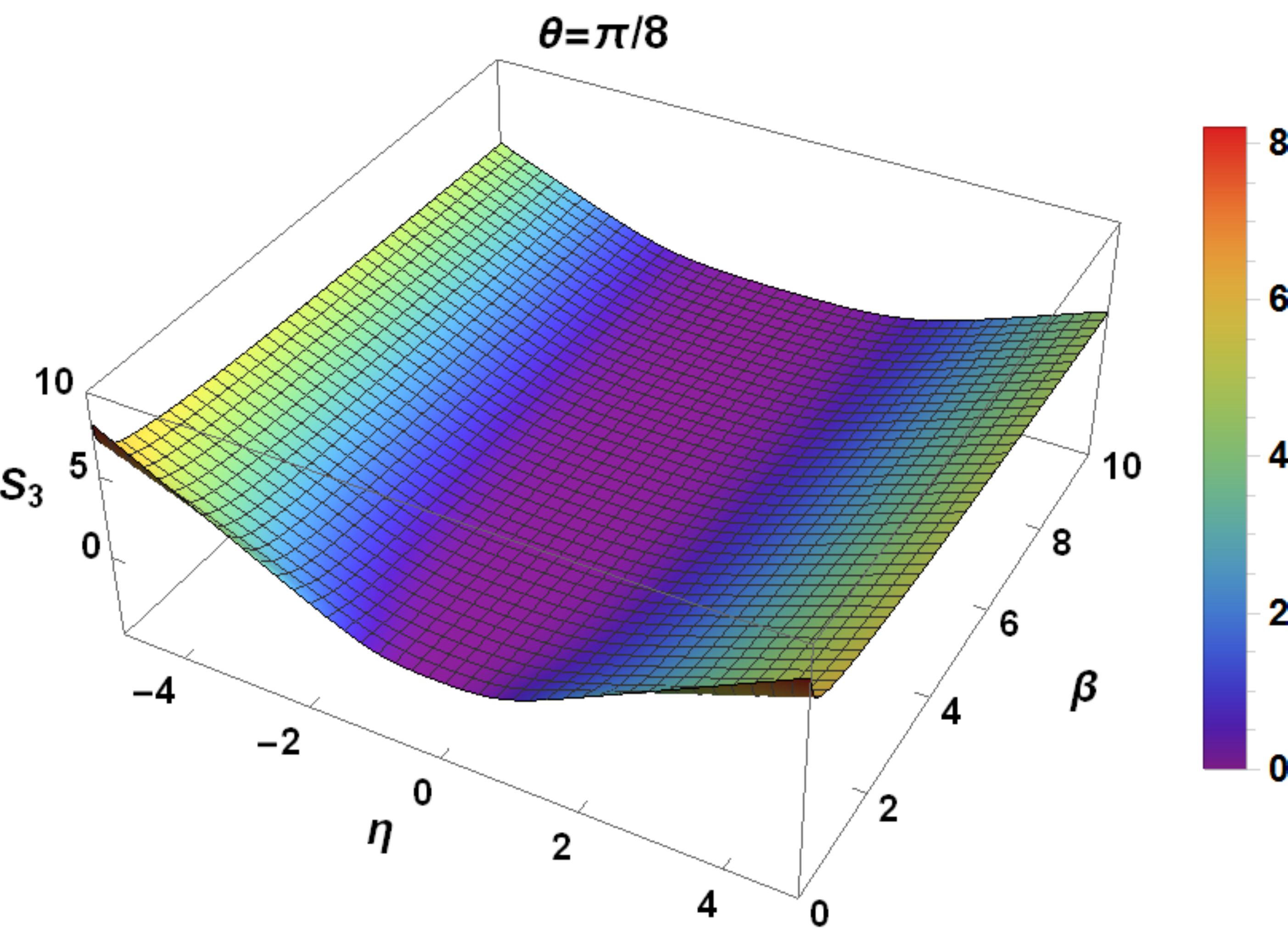}
\caption{\textsf{R\'{e}nyi entropy $S_3$ versus the coupling parameter $\protect\eta$ and the
temperature  $\beta$ for fixed values of the mixing angle $\theta=\frac{\pi}{2}, \frac{\pi}{3}, \frac{\pi}{4}, \frac{\pi}{8}$. }}
\label{R52-theta}
\end{figure}

Figure \ref{R52-eta} shows the R\'{e}nyi entropy $S_3$ as function of the temperature $\beta$ and the
mixing angle $\protect\theta$ for  four values of the coupling parameter $\eta=1,2,3,4$.
We observe that there are periodicity with respect to $\theta$ such that the same behavior
repeats in $[0,\pi]$ and $[\pi, 2\pi]$.
It is clearly seen
that $S_3$ is maximal at high temperature while it is minimal for low temperature.
As long as $\eta$ is increased, we notice that $S_3$ increases rapidly to reach the maxima values
as shown for the case $\eta=4$.
In Figure \ref{R52-theta}, we present the
R\'{e}nyi entropy $S_3$ as function of the coupling parameter $\protect\eta$ and the
temperature  $\beta$ for fixed values of the mixing angle $\theta=\frac{\pi}{2}, \frac{\pi}{3}, \frac{\pi}{4}, \frac{\pi}{8}$.
We observe that $S_3$ shows a symmetric behavior with  respect to the value $\eta=0$ and decreases as long as
$\theta$  decreased from   $\frac{\pi}{2}$ to $\frac{\pi}{8}$.
%
We conclude that the R\'{e}nyi entropy
can be controlled and adjusted by different parameters to  extract some information about  our system.
This is clearly seen from different configurations chosen to obtain 
such plots in many shapes of Figures \ref{R52-beta},\ref{R52-eta},\ref{R52-theta}.


\section{von Neumann entropy}


To accomplish our study about some entropies, we establish a relation between
the von Neumann entropy and the purity function \eqref{35} of our system. 
%
%
%
This can be worked out to end up with the 
expression
\begin{equation}
S_{1}=S_{vN}\left( \eta,\theta;\beta\right) =-\ln\left( \frac{2P }{1+P }%
\right) -\frac{1-P }{2P }\ln\frac{1-P }{1+P }
\end{equation}
which corresponds to the case $q=1$ in general form of
the R\'{e}nyi entropy as seen before. 
We notice  that, such entropy 
is
also a function of three physical parameters $\eta, \theta$ and $\beta$ characterizing
our system.
%
%

\begin{figure}[H]
\centering  \includegraphics[width=8cm,  height=5.1cm]{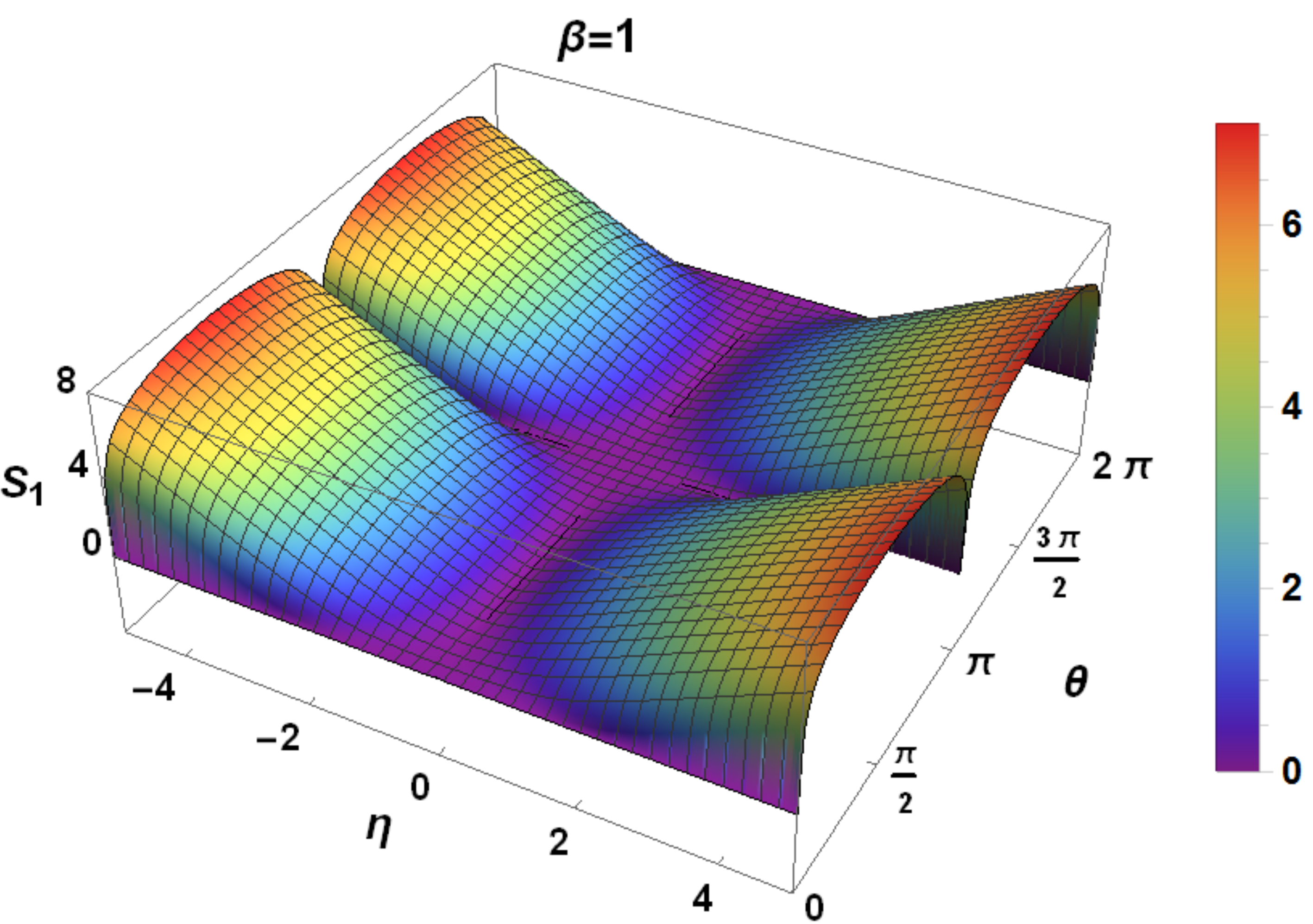}  %
\includegraphics[width=8cm,  height=5.1cm]{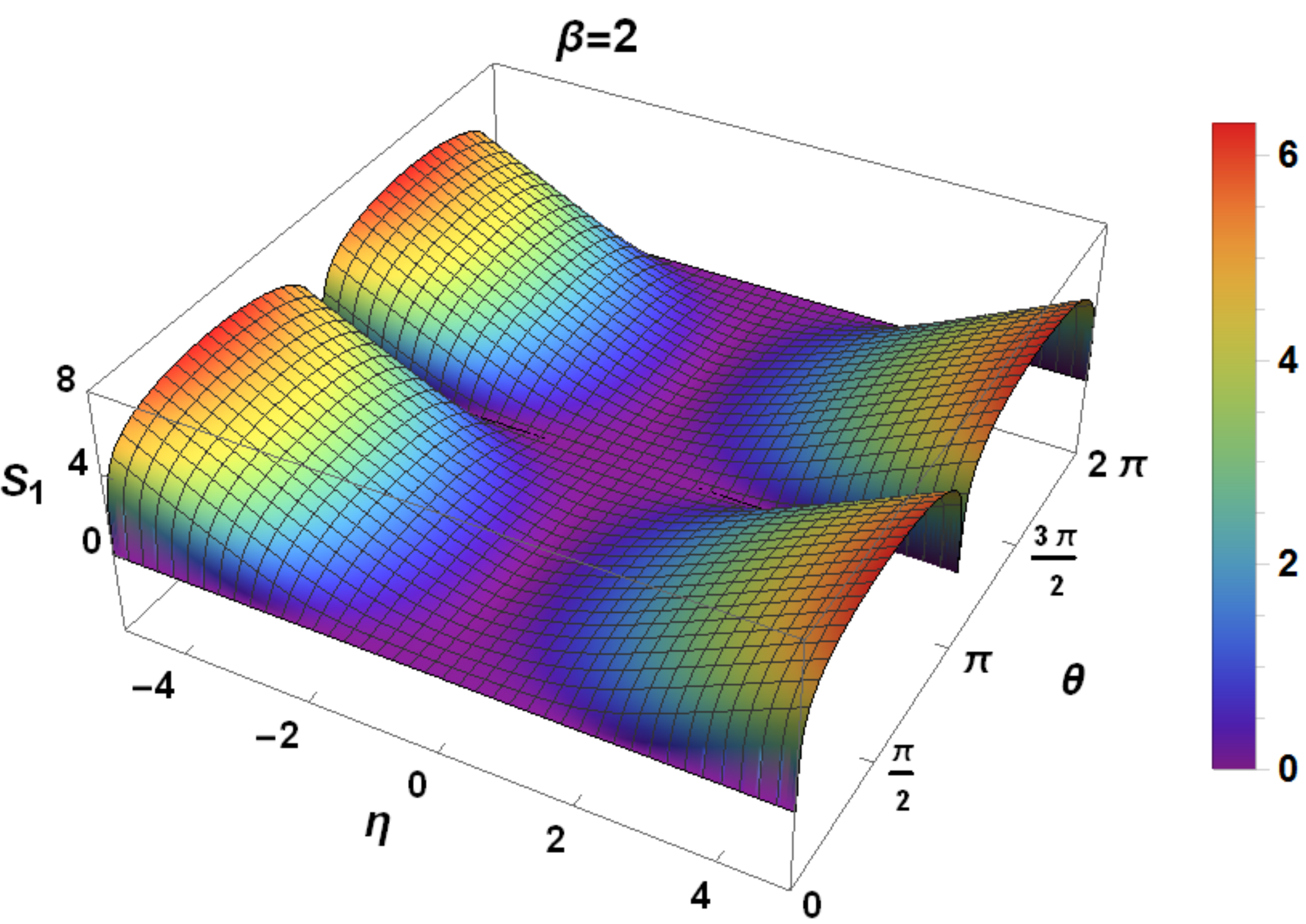}  %
\includegraphics[width=8cm,  height=5.1cm]{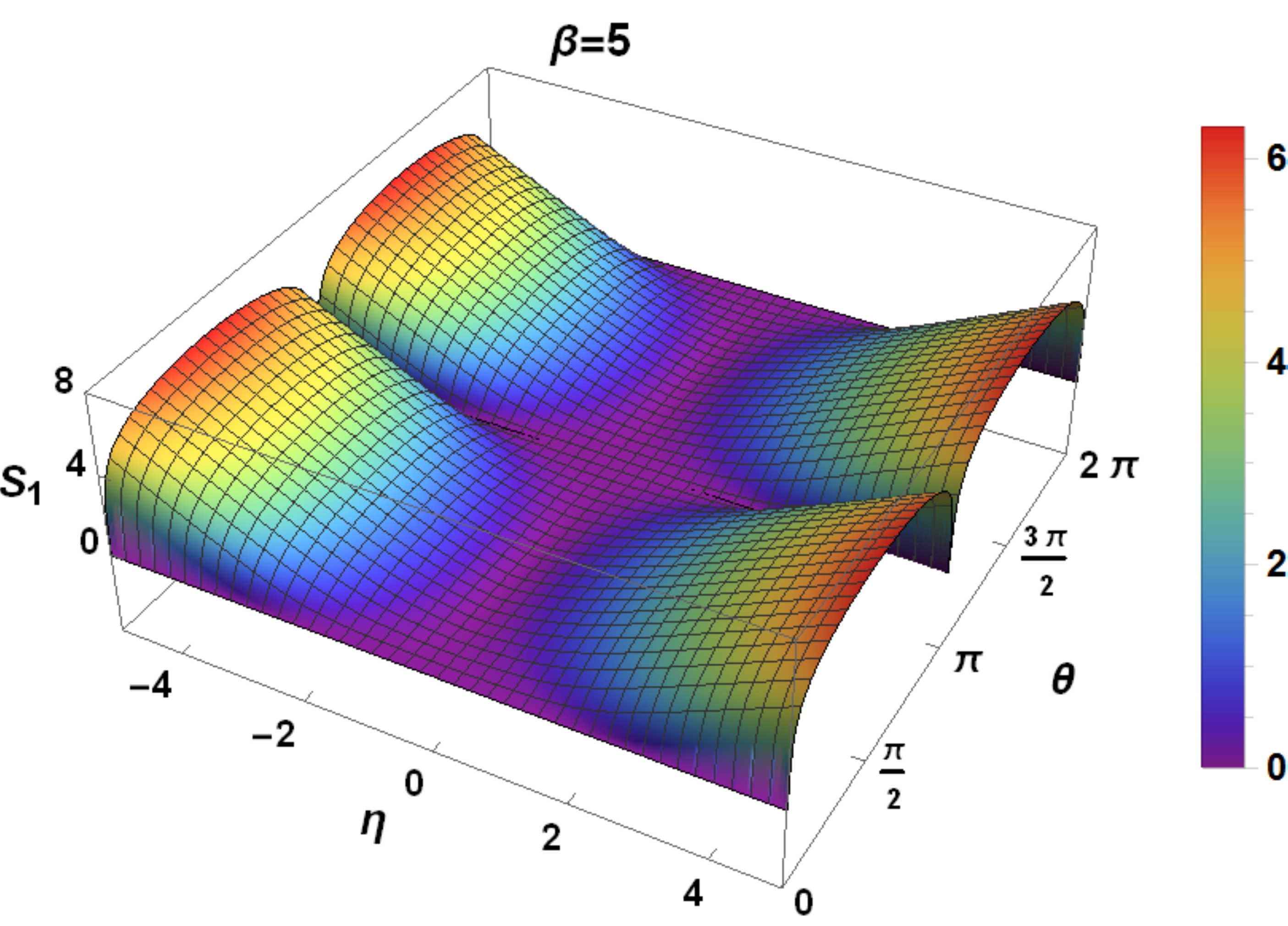}  %
\includegraphics[width=8cm,  height=5.1cm]{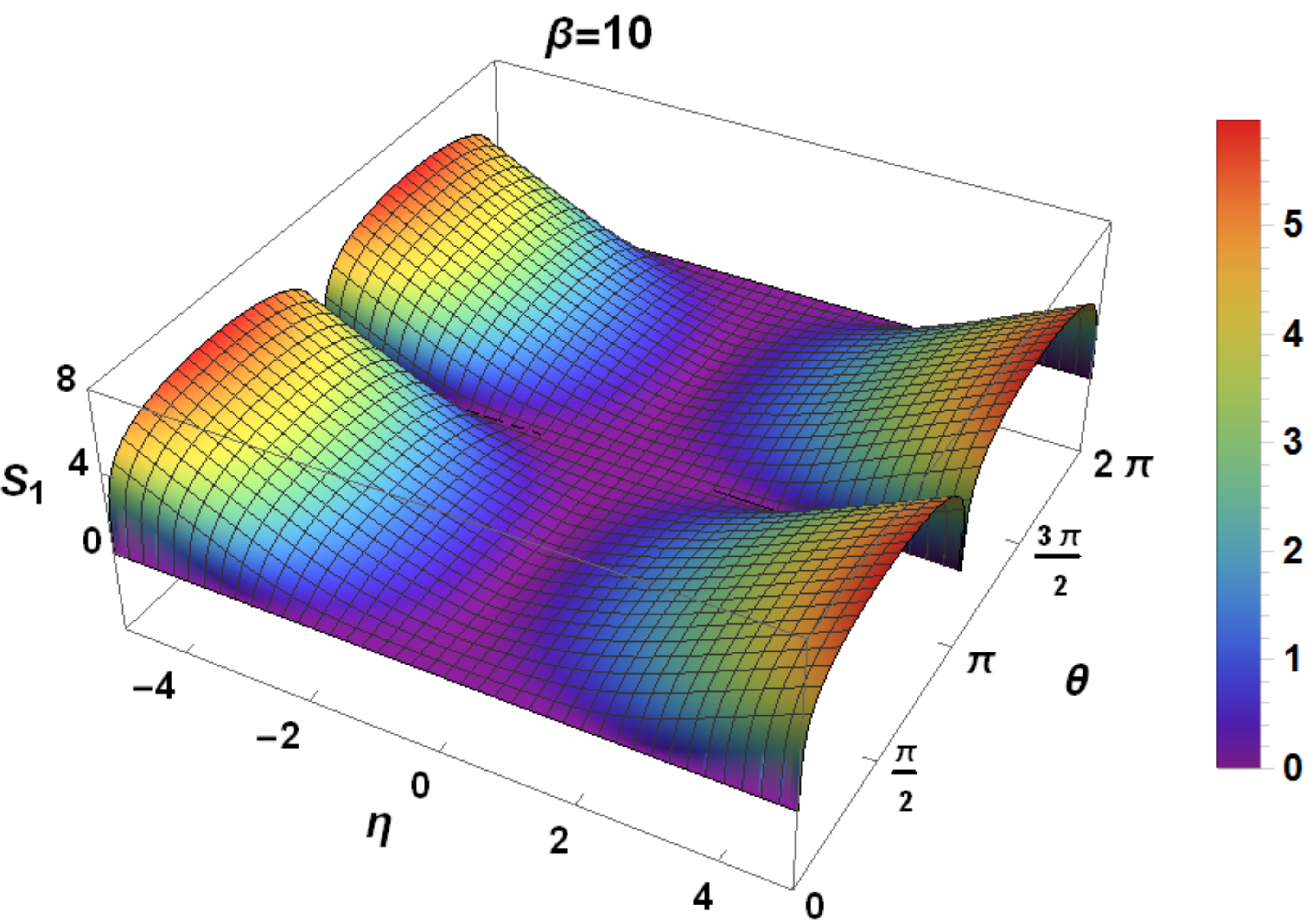}
\caption{\textsf{von Neumann entropy versus the coupling parameter $\protect\eta$ and the
mixing angle $\protect\theta$ for  fixed values of the temperature $\protect\beta=1,2,5,10$. }}
\label{R58-beta} 
\end{figure}

\begin{figure}[H]
\centering  \includegraphics[width=8cm,  height=5.1cm]{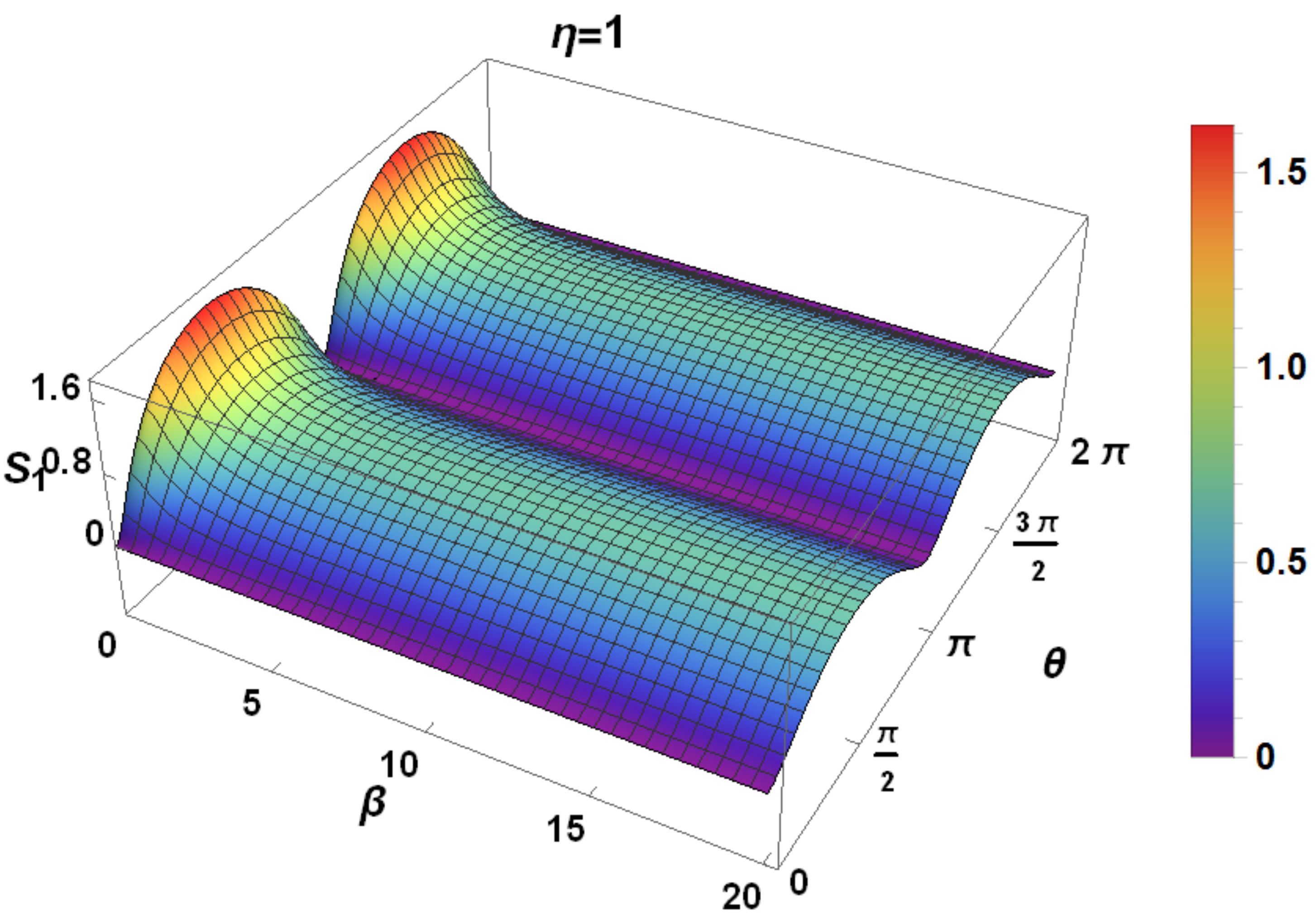}  %
\includegraphics[width=8cm,  height=5.1cm]{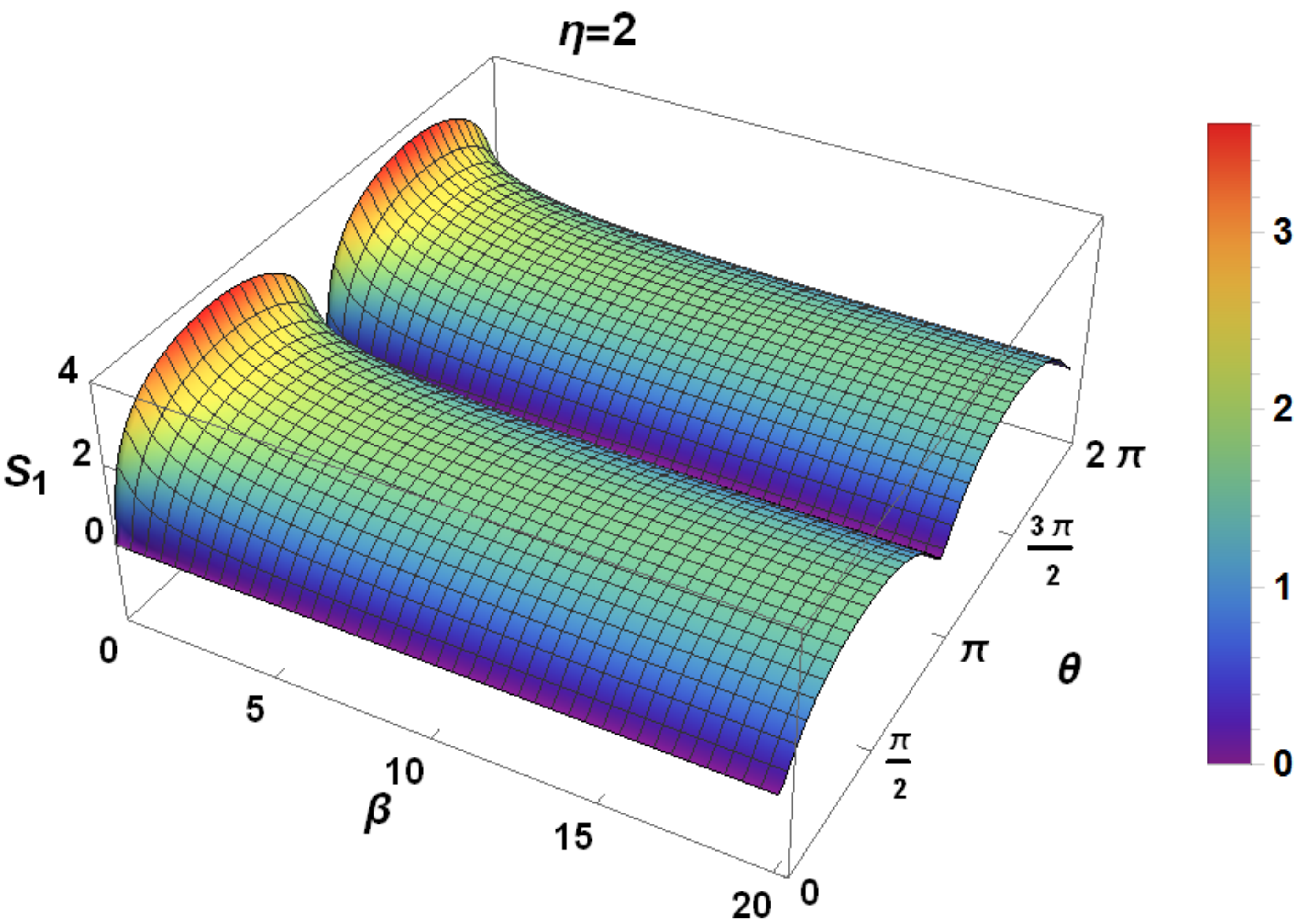}  %
\includegraphics[width=8cm,  height=5.1cm]{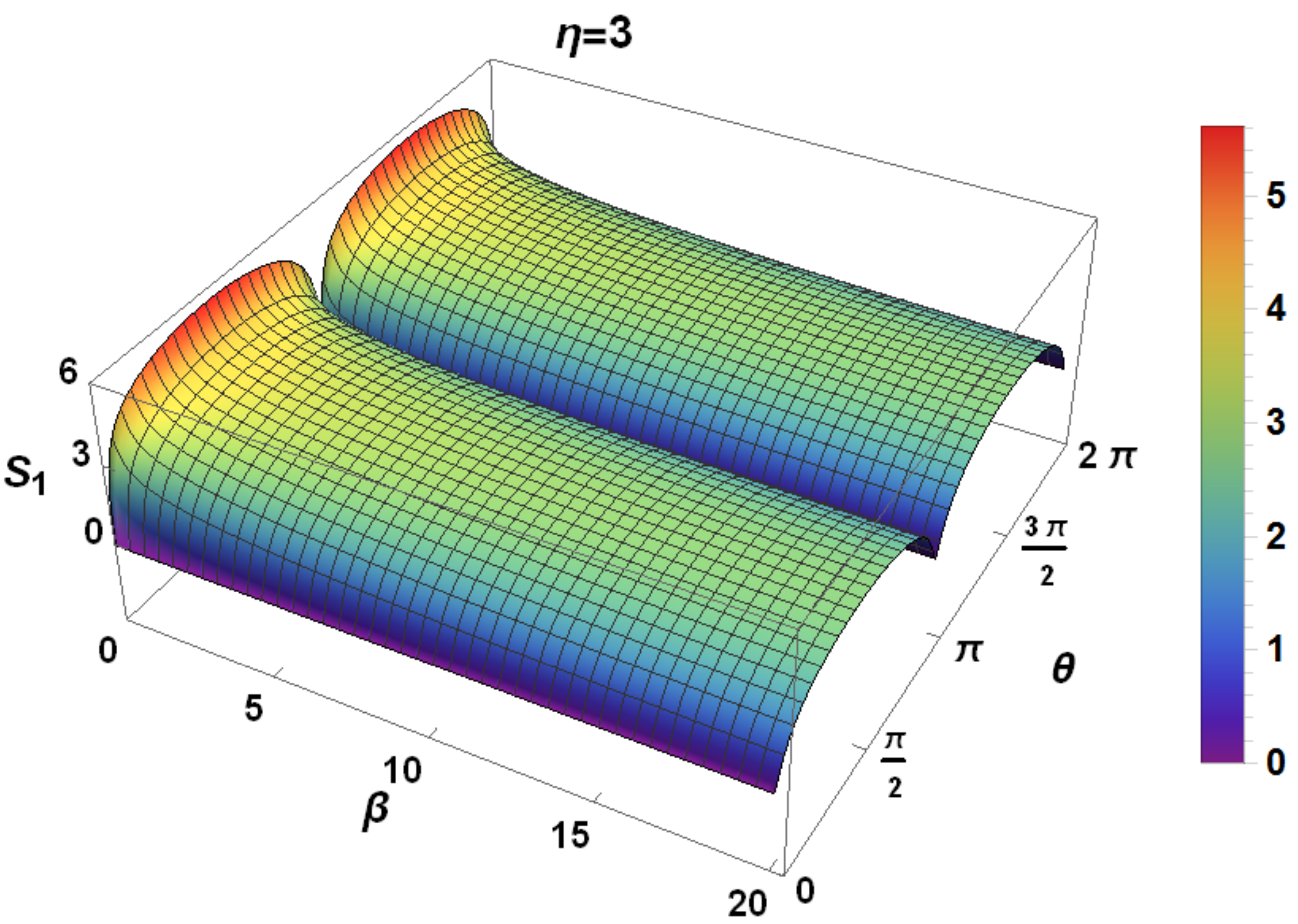}  %
\includegraphics[width=8cm,  height=5.1cm]{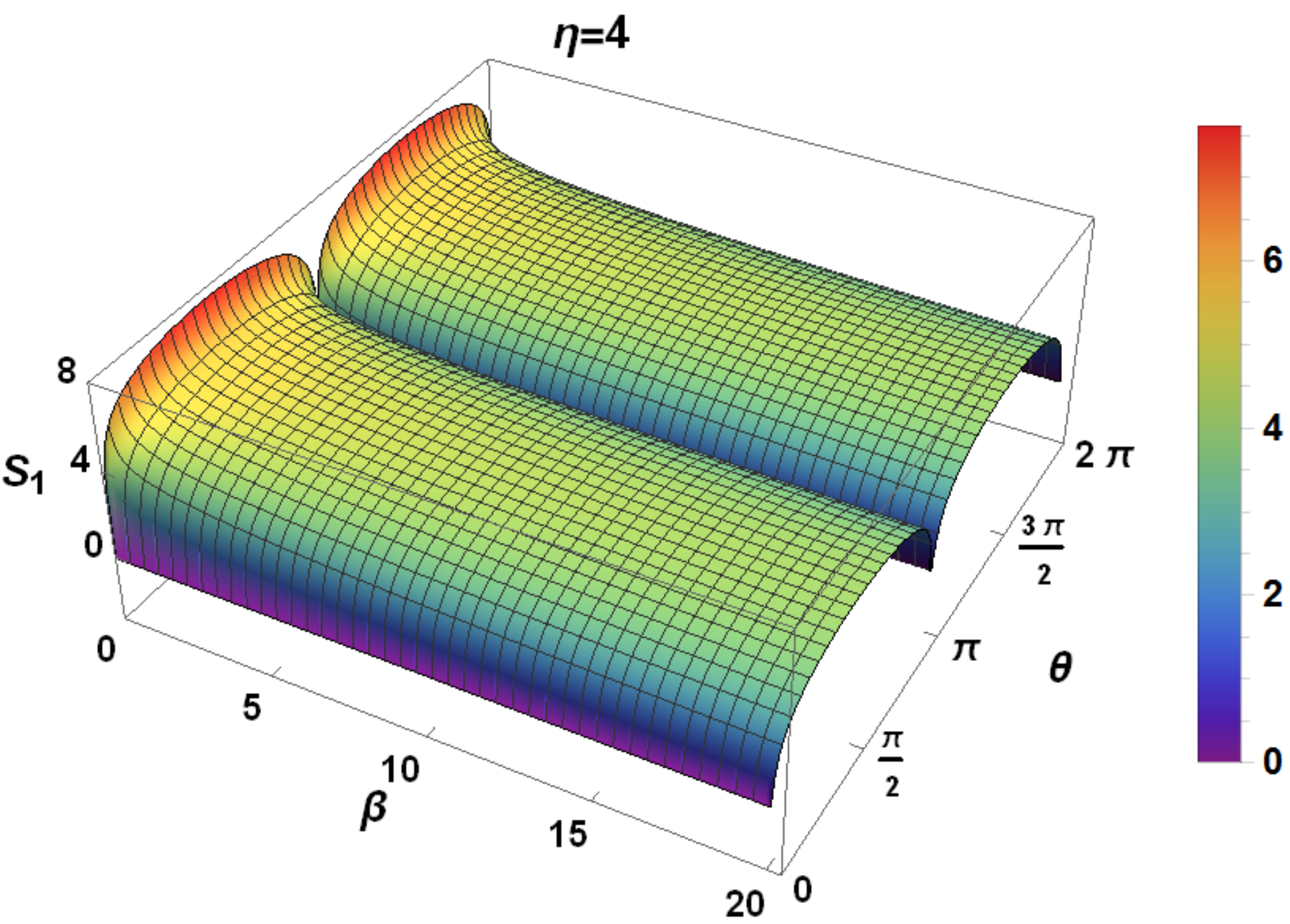}
\caption{\textsf{von Neumann entropy versus the temperature $\beta$ and the
mixing angle $\protect\theta$ for  fixed values of the coupling parameter $\eta=1,2,3,4$. }}
\label{R58-eta} 
\end{figure}

\begin{figure}[H]
\centering  \includegraphics[width=8cm,  height=5.1cm]{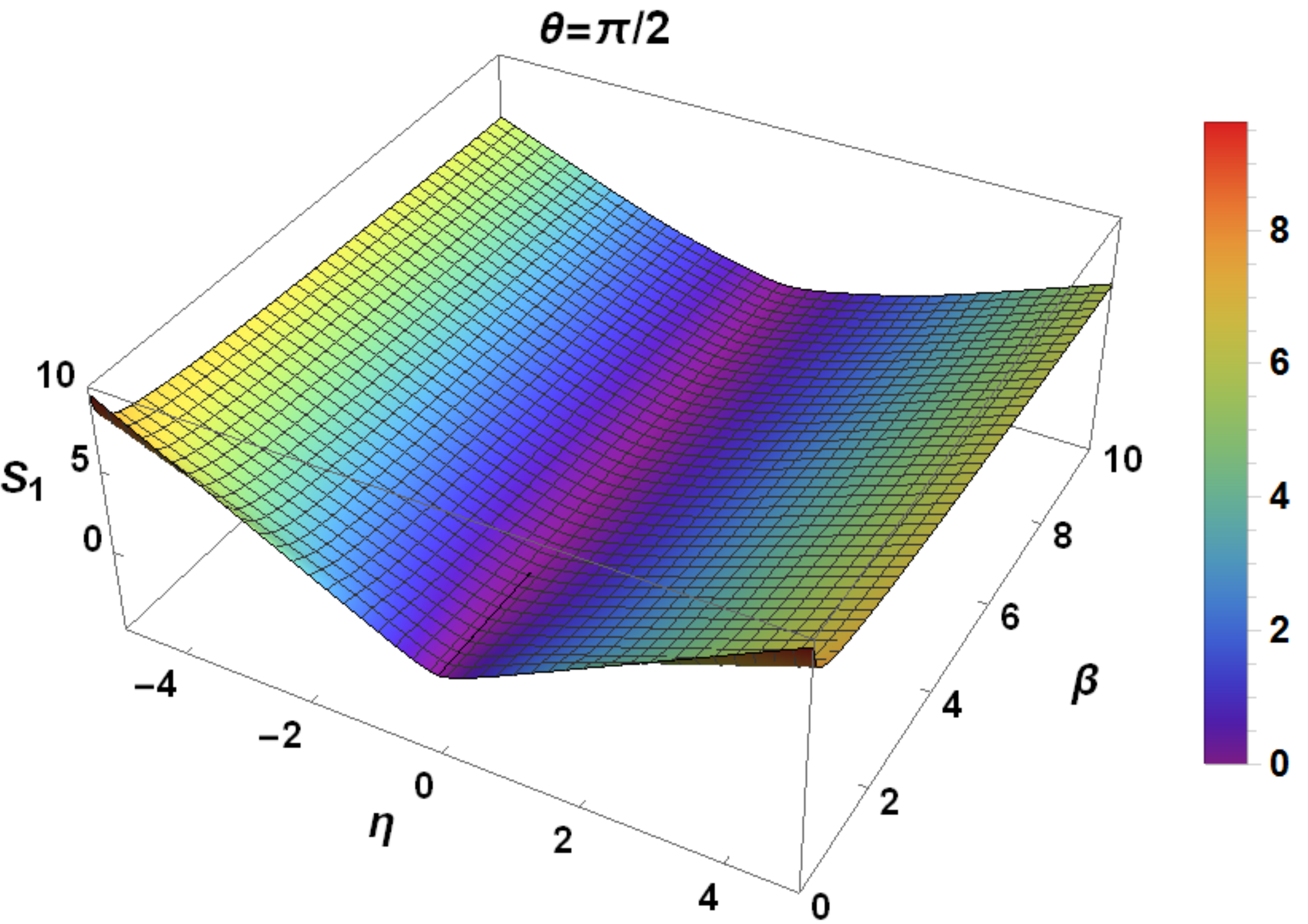}  %
\includegraphics[width=8cm,  height=5.1cm]{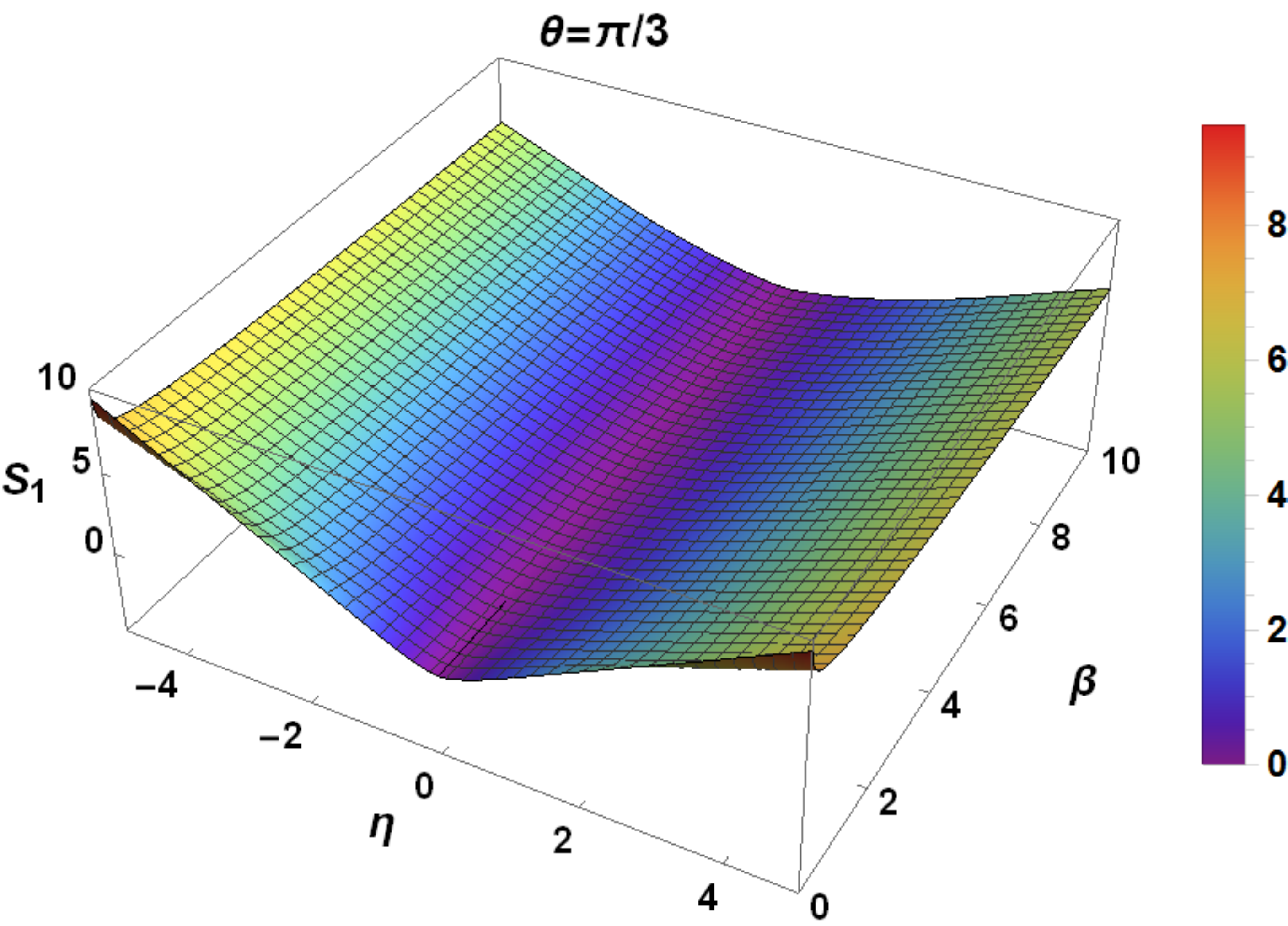}  %
\includegraphics[width=8cm,  height=5.1cm]{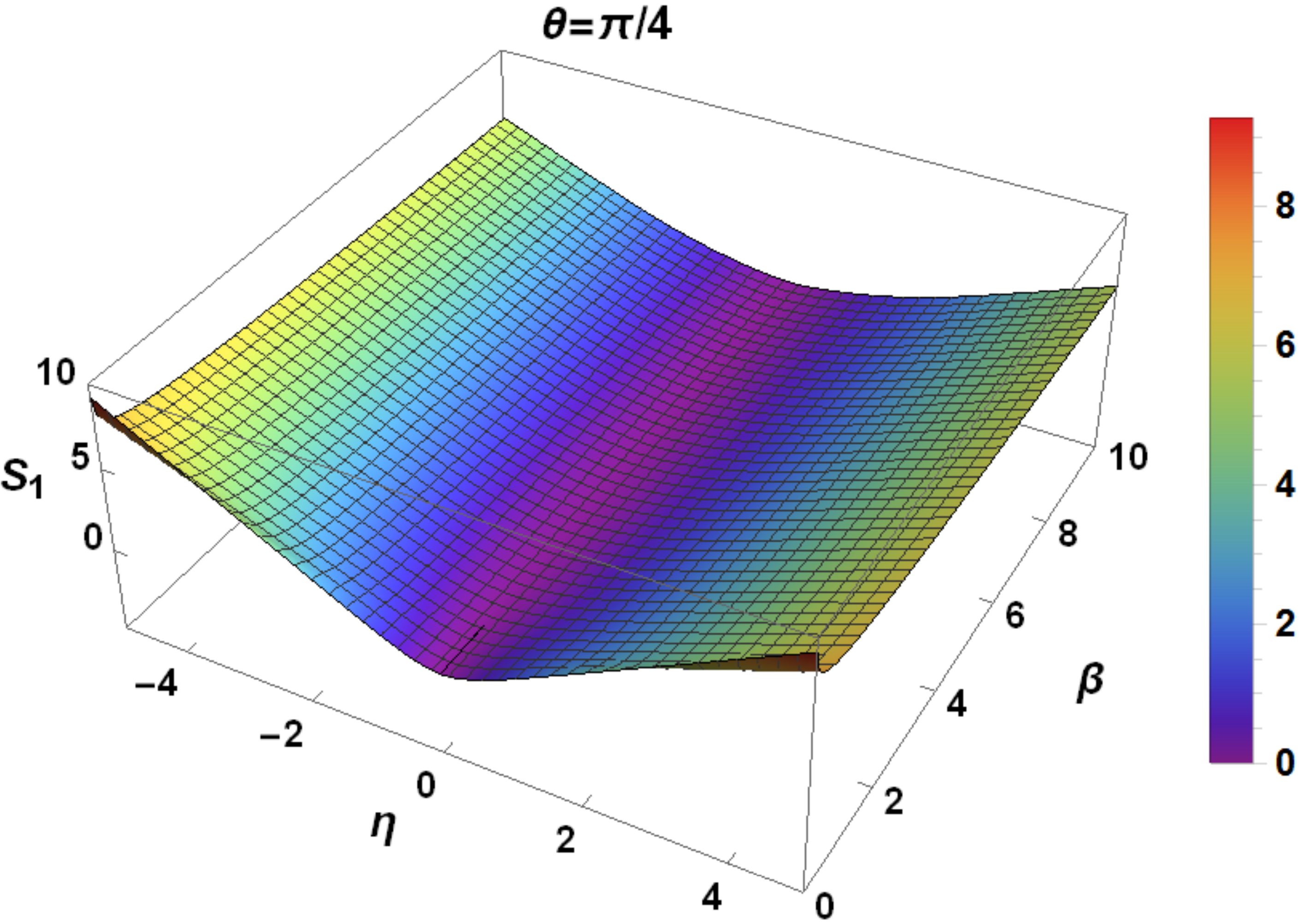}  %
\includegraphics[width=8cm,  height=5.1cm]{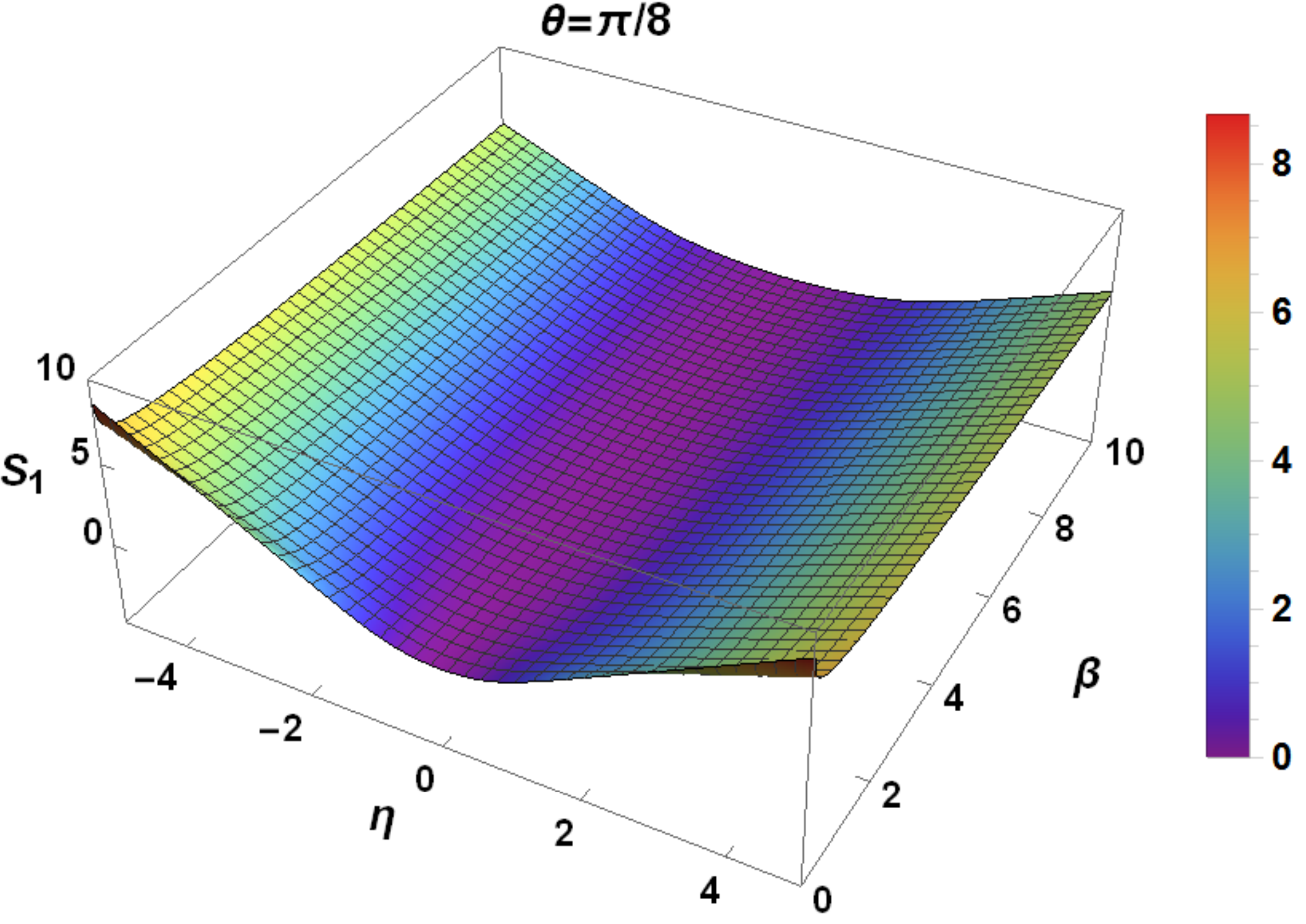}
\caption{\textsf{ von Neumann entropy versus the coupling parameter $\protect\eta$ and the
temperature  $\beta$ for fixed values of the mixing angle $\theta=\frac{\pi}{2}, \frac{\pi}{3}, \frac{\pi}{4}, \frac{\pi}{8}$. }}
\label{R58-theta}
\end{figure}

Figure \ref{R58-beta}  shows the von Neumann entropy $S_1$ as function of the coupling parameter 
$\protect\eta$ and the
mixing angle $\protect\theta$ for  fixed values of the temperature $\protect\beta=1,2,5,10$.
Figure \ref{R58-eta}  presents $S_1$ as function of  the temperature $\beta$ and the
mixing angle $\protect\theta$ for  fixed values of the coupling parameter $\eta=1,2,3,4$.
Figure \ref{R58-theta} shows $S_1$ as function
of the the coupling parameter $\protect\eta$ and the
temperature  $\beta$ for fixed values of the mixing angle $\theta=\frac{\pi}{2}, \frac{\pi}{3}, 
\frac{\pi}{4}, \frac{\pi}{8}$. Compared to those of the R\'{e}nyi entropy $S_3$, such Figures 
present some some similarities and differences.

To close our study it is interesting
to summarize in three different tables below the most interesting form can
be taken by the von Neumann entropy $S_1$
for particular values
of the coupling parameter and mixing angle as well as low and high temperature regimes.
Indeed, we start by analyzing
two situations with respect to the strength of the
coupling parameter $\eta$, which will allow us to underline the behavior of our system. 
We start with the weak coupling that is characterized by taking the limit  $%
C_{3}\longrightarrow0$ where the angle $\theta\longrightarrow\theta_{w}$ and
the coupling $\eta\longrightarrow\eta_{w}$. In this case, 
\eqref{theta} and \eqref{eta} reduce to the following quantities
\begin{equation}
\theta_{w}=0, \qquad 
e^{2\eta_{w}}=\frac{1}{\mu^{2}}\sqrt{\frac{C_{1}}{C_{2}}}.
\end{equation}
Now we consider the strong coupling limit and derive in the beginning the corresponding physical parameters.
In doing so, we notice that if the limit  $C_{3}\longrightarrow2%
\sqrt{C_{1}C_{2}}$ is required  then one can end up with the limits 
\begin{eqnarray}
&& \tan\theta_{s}\longrightarrow \frac{2\sqrt{C_{1}C_{2}}}{\mu^{2}C_{2}-%
\frac{C_{1}}{\mu^{2}}}\longrightarrow0 \\
&& \eta\longrightarrow\eta_{s}=+\infty, \qquad k\longrightarrow0^{+}
\end{eqnarray}
giving rise to the results
\begin{eqnarray}
ke^{2\eta_{s}} \longrightarrow\frac{C_{1}}{\mu^{2}}+\mu^{2}C_{2}, \qquad
\theta_{s} =\tan^{-1}\left( \frac{2\sqrt{C_{1}C_{2}}}{\mu^{2}C_{2}-\frac{%
C_{1}}{\mu^{2}}}\right).
\end{eqnarray}
Combining all to write the  von Neumann entropies describing both limiting cases 
in Table {\color{red}1}, which is either zero or infinity.

\begin{center}
\begin{tabular}{||c|c|c|c||}
\hline
Coupling & Angle & Purity & von Neumann entropy \\[0.5ex] \hline\hline
$\eta$ & $\theta$ & $P\left( \beta\right) $ & $S_{vN}\left( \beta\right)$ \\
\hline
$\eta_{w}$ & $\theta_{w}$ & $1$ & $0$ \\ \hline
$\eta_{s}$ & $\theta_{s}$ & $0$ & $\infty$ \\[1ex] \hline
\end{tabular}
\end{center}
{\sf
Table 1: The von Neumann entropy $S_1$ as function of temperature $\beta$ for strong $\eta=\eta_s$ and weak 
$\eta=\eta_w$ coupling.
}

The last situation is related to the nature of our system, 
which is equivalent to require that both of  harmonic oscillators have the same
mass $m_{1}=m_{2}$ and
frequency $C_{1}=C_{2}$. Thus from \eqref{theta} and \eqref{eta}, we end up with the constraint
$\theta\longrightarrow\frac{\pi}{2}$ and $\eta\longrightarrow \eta_{id}$
with
\begin{equation}
e^{2\eta_{id}}=\sqrt{\frac{C_{1}+\frac{C_{3}}{2}}{C_{1}-\frac{C_{3}}{2}}}.
\end{equation}
The corresponding entropies can be summarized as function of 
the temperature
 Table {\color{red} 2} and function of the coupling parameter $\eta_{id}$ (identical masses $m_1=m_2$) 
 Table {\color{red} 3}. It is clearly see that in all cases we have different forms of the  von Neumann
 entropies, which can be simplified by replacing the purity function by their forms under the conditions
 taken into consideration.
\vspace{2.5mm}
\begin{center}
\begin{tabular}{||c|c|c|c||}
\hline
Coupling & Angle & Purity & von Neumann entropy \\[0.5ex] \hline\hline
$\eta$ & $\theta$ & $P\left(\beta\right)$ & $S_{vN}\left(\beta\right)$ \\
\hline
$\eta_{id}$ & $\frac{\pi}{2}$ & $\frac{2\sqrt{\tanh\left( {\hbar}%
\sqrt{\frac{k_{id}}{m}}e^{\eta_{id}}\beta\right) \tanh\left( {\hbar}%
\sqrt{\frac{k_{id}}{m}}e^{-\eta_{id}}\beta\right) }}{e^{\eta_{id}}\tanh%
\left( {\hbar}\sqrt{\frac{k_{id}}{m}}e^{\eta_{id}}\beta\right)
+e^{-\eta_{id}}\tanh\left( {\hbar}\sqrt{\frac{k_{id}}{m}}%
e^{-\eta_{id}}\beta\right) }$ & $-\ln\left( \frac{2P_{id}}{1+P_{id}}\right) -%
\frac{1-P_{id}}{2P_{id}}\ln\frac{1-P_{id}}{1+P_{id}}$ \\[1ex] \hline
\end{tabular}
\end{center}
{
\sf Table 2: The von Neumann entropy $S_1$ as function of temperature $\beta$ for identical particules $\eta=\eta_{id}$
and mixing angle $\theta=\frac{\pi}{2}$.}

\vspace{2.5mm}
\begin{center}
\begin{tabular}{||c|c|c|c||}
\hline
Temperature & Angle & Purity & von Neumann entropy \\[0.5ex] \hline\hline
$\beta$ & $\theta$ & $P\left( \eta_{id}\right) $ & $S_{vN}\left( \eta
_{id}\right) $ \\ \hline
$\infty$ & $\theta_{id}=\frac{\pi}{2}$ & $\frac{1}{\cosh\left( \eta
_{id}\right) }$ & $2\left( 1-\sinh^{2}\left( \frac{\eta_{id}}{2}\right)
\right) \ln\left( \cosh\left( \frac{\eta_{id}}{2}\right) \right)
-\sinh^{2}\left( \frac{\eta_{id}}{2}\right) \ln\left( \sinh^{2}\left( \frac{%
\eta_{id}}{2}\right) \right) $ \\ \hline
$0$ & $\theta_{id}=\frac{\pi}{2}$ & $\frac{1}{\cosh\left( 2\eta_{id}\right) }
$ & $2\left( 1-\sinh^{2}\left( \eta_{id}\right) \right) \ln\left(
\cosh\left( \eta_{id}\right) \right) -\sinh^{2}\left( \eta_{id}\right)
\ln\left( \sinh^{2}\left( \eta_{id}\right) \right) $ \\[1ex] \hline
\end{tabular}
\end{center}
{\sf
Table 3: The von Neumann entropy $S_1$ as function of coupling parameter $\eta=\eta_{id}$ with
mixing angle 
$\theta=\frac{\pi}{2}$ for high and low temperature.}

\section{Conclusion}


We have studied two interesting entropies 
for a system
of two coupled harmonic oscillators by using the path integral
mechanism. In doing so, we have involved a global propagator
based on temperature evolution of our system. Considering
a unitary transformation we were able to explicitly obtain 
the reduced density matrix and therefore the thermal wavefunction describing the whole
spectrum of our system. These allowed us to derive the purity function
characterizing the entanglement of our system in terms of temperature 
and coupling parameter \cite{Merdaci18}.

We have used our previous results obtained in \cite{Merdaci18} to build in the first stage the R\'{e}nyi entropies $S_q$
for all parameter $q>1$. To illustrate such study we have focused on
$q=3$ and presented different plots showing the particularities of the entropy $S_3$.
Subsequently, we have determined the von Neumann entropy $S_1$, which corresponds to
the limiting case $q\lga 1$ of the  R\'{e}nyi entropies. We numerically analyzed
$S_1$ by offering some plots under some choice of the coupling parameter, rotating angle
and temperature.
For its relevance
we have considered particular cases and derived the corresponding von Neumann entropies. For this, we have
gave three different tables showing the values can be taken by $S_1$ according to the
nature of our system as well as the temperature regime.


\section*{Acknowledgments}


We thank Youness Zahidi for his numerical help.
The authors acknowledge the financial support from the Deanship
of Scientific Research (DSR) of King Faisal University. 
 The present work was done under Project Number `180118',
Purity Temperature Dependent for two Coupled Harmonic
Oscillators. 






\begin{thebibliography}{99}

\bibitem{Shannon}
C. E. Shannon, “A Mathematical Theory of Communication,” Bell System Technical Journal, Vol. 27, 1948, pp.
379-423 and 623-656. 


\bibitem{Renyi2} A. Renyi, ''On measures of entropy and information``, 
Proc. Fourth Berkeley Symp. on Math. Statist. Prob., Vol. 1 (Univ. of Calif. Press, 1961), 547.



\bibitem{21} H. Li and F. D. M. Haldane, Phys. Rev. Lett. 101, 010504
(2008), 0805.0332.

\bibitem{22} M. Headrick, Phys. Rev. D 82, 126010 (2010), 1006.0047.


\bibitem{Merdaci18} A. Merdaci, A. Jellal, A. Al Sawalha and A. Bahaoui, J. Stat. Mech. (2018) 093101.



\bibitem{jellalstat} A. Jellal,  F. Madouri and A. Merdaci, J. Stat. Mech.
(2011) P09015.

\bibitem{jellal} A. Jellal, E.H. El Kinani and M. Schreiber, Int. J. Mod.
Phys. A {20} (2005) 1515.

\bibitem{Kosztin} I. Kosztin, B. Faber and K. Schulten, 
Am. J. Phys. 64 (1996)  633.


\bibitem{rossi} M. Rossi, M. Nava, L. Reatto and D.E. Galli,
 J. Chem. Phys {131} (2009) 154108. 

 
\bibitem{Kleinert} H. Kleinert, \textit{"Path Integrals in Quantum
Mechanics, Statistics, Polymer Physics, and Financial Markets``} (World
Scientific, Singapore 2009). 

\bibitem{Renyi} A. R\`enyi, \textit{''Probability Theory``} (North
Holland, Amsterdam, 1970).

\bibitem{Bastiaans} M. J. Bastiaans, J. Opt. Soc. Am. {1} (1984) 711;
ibid. {3} (1986) 1243.

\bibitem{Tsallis} C. Tsallis, J. Stat. Phys. {52} (1988) 479.



\bibitem{Adesso2} G. Adesso, A. Serafini and F. Illuminati, Phys. Rev. A {70} (2004) 022318.

\bibitem{Pipek} J. Pipek and I. Nagy, Phys. Rev. A {79} (2009) 052501.


\bibitem{Adesso} G. Adesso, A. Serafini and F. Illuminati, Open Syst. Inf.
Dyn. {12} (2005) 189.

\end{thebibliography}
\end{document}